\documentclass[11pt]{article}
\usepackage{amsmath,amsthm,bm, amscd}
\usepackage{fullpage}
\usepackage[nofullpage,full,titlepage,nousetoc,nouselot,nouselof,hylinks,final]{boaz}

\usepackage{graphicx}

\newcommand{\supp}{\mathsf{supp}}

\theoremstyle{definition}

\newcommand{\Raz}{\mathsf{Raz}}
\newcommand{\Slice}{\mathsf{Slice}}  

\newcommand{\Ext}{\mathsf{Ext}}
\newcommand{\Edit}{\mathsf{Edit}}
\newcommand{\SAuth}{\mathsf{SAuth}}
\newcommand{\Extract}{\mathsf{Extract}}
\newcommand{\ExtractH}{\mathsf{ExtractH}}
\newcommand{\NExtract}{\mathsf{NExtract}}
\newcommand{\Auth}{\mathsf{Auth}}
\newcommand{\NAuth}{\mathsf{NAuth}}
\newcommand{\SRGExt}{\mathsf{SRGExt}}
\newcommand{\Pre}{\mathsf{Pre}}
\newcommand{\EditDis}{\mathsf{EditDis}}
\newcommand{\TExt}{\mathsf{TExt}}
\newcommand{\zuc}{{\rm Zuc}}

\newcommand{\BI}{\begin{itemize}}
\newcommand{\EI}{\end{itemize}}
\newcommand{\BE}{\begin{enumerate}}
\newcommand{\EE}{\end{enumerate}}

\newtheorem{thm}{Theorem}      
\newcommand{\BT}{\begin{theorem}}   \newcommand{\ET}{\end{theorem}}
\newcommand{\BD}{\begin{definition}}   \newcommand{\ED}{\end{definition}}
\newcommand{\BCR}{\begin{corollary}} \newcommand{\ECR}{\end{corollary}}
\newtheorem{constr}[thm]{Construction}
\newcommand{\BCT}{\begin{constr}} \newcommand{\ECT}{\end{constr}}
\newcommand{\BL}{\begin{lemma}}   \newcommand{\EL}{\end{lemma}}

\newcommand{\BP}{\begin{proposition}}   \newcommand{\EP}{\end{proposition}}
\newcommand{\BCM}{\begin{claim}}   \newcommand{\ECM}{\end{claim}}
\newcommand{\BF}{\begin{fact}}   \newcommand{\EF}{\end{fact}}
\newcommand{\BA}{\begin{assumption}}   \newcommand{\EA}{\end{assumption}}

\begin{document}

\begin{titlepage}
\def\thepage{}

\date{}
\title{On the Problem of Local Randomness in Privacy Amplification with an Active Adversary}

\author{
Xin Li \thanks{Supported in part by NSF Grant CCF-0634811, NSF grant CCF-0916160 and
THECB ARP Grant 003658-0113-2007. }\\
University of Texas at Austin\\
lixints@cs.utexas.edu\\
}

\maketitle \thispagestyle{empty}

\begin{abstract}
We study the problem of privacy amplification with an active adversary in the information theoretic setting. In this setting, two parties Alice and Bob start out with a shared $n$-bit weak random string $W$, and try to agree on a secret random key $R$ over a public channel fully controlled by an active and unbounded adversary. Typical assumptions are that these two parties have access to local private uniform random bits. In this paper we seek to minimize the requirements on the local randomness used by the two parties. 

We make two improvements over previous results. First, we reduce the number of random bits needed for each party to $\Theta(\ell+\log n)$, where $\ell$ is the security parameter, as long as $W$ has min-entropy $n^{\Omega(1)}$. Previously, the best known result needs to use $\Theta((\ell+\log n)\log n)$ bits. Our result is also asymptotically optimal. Second, we generalize the problem to the case where the two parties only have local weak random sources instead of truly uniform random bits. We show that when each party has a local weak random source with min-entropy $> n/2$, there is an efficient privacy amplification protocol that works nearly as good as if the two parties have access to local uniform random bits. Next, in the case where each party only has a weak random source with arbitrarily linear min-entropy, we give an efficient privacy amplification protocol where we can achieve security parameter up to $\Omega(\log k)$. Our results give the first protocols that achieve privacy amplification when each party only has access to a local weak random source. 
\end{abstract}
\end{titlepage}

\section{Introduction}
The problem of privacy amplification, introduced in a paper by Bennett, Brassard, and Robert \cite{BennettBR88}, is a fundamental problem in symmetric key cryptography. In symmetric key cryptography, two parties (Alice and Bob) share an $n$-bit secret $W$ and they wish to communicate over a public channel controlled by an adversary Eve. The goal is to make the communication private and authentic. While this problem is well studied in cryptography, standard solutions require that $W$ is perfectly uniform. The problem of privacy amplification, on the other hand, assumes that $W$ is not uniform and is some arbitrary probability distribution. As in \cite{BennetBCM95}, in this paper we only require that $W$ has a certain amount of entropy\footnote{We use standard notions of min-entropy.}. The goal now is for Alice and Bob to agree on a string $R$, that is nearly uniform and secret, so that standard solutions can be evoked. This problem is natural and important for the following reasons.

First, in practice many secrets are not uniformly distributed. In fact, despite the fact that randomness plays a very important role in computer science, it is not clear how to obtain uniform random bits in the real world. Many random like behaviors, such as weather patterns, biometrics, human-memorable passphrases and market prices are highly biased distributions. Second, even if the two parties manage to acquire a shared uniform random string, it could be compromised to some extent. For example the adversary can use side-channel attacks against the hardware. In this case it is often reasonable to assume that the adversary does not learn the entire information of the secret (otherwise security is completely lost), and that the secret has some amount of entropy left. The problem of privacy amplification thus arises naturally in either of these cases.

To study this problem, it is necessary to model the adversary. One possibility is to assume that the adversary is computationally bounded, so that cryptographic assumptions and primitives can be used \cite{BMP00, KOY01, CHK05, GL06}. On the other hand, in this paper we make no computational assumptions and instead assume that the adversary is computationally unbounded. Thus here we are studying the problem in the \emph{information theoretic setting}.

We can also model the adversary as being passive or active. A passive adversary just listens to the communication over the channel and does not change anything. An active adversary, on the other hand, can modify the communication in arbitrary ways. We note that in the case of a passive adversary, strong extractors \cite{NisanZ96} can be used to give an optimal solution to this problem. However in this paper, we assume that the adversary is active.

With an active and computationally unbounded adversary, the problem becomes considerably harder. To see this, note that now the adversary can modify the messages sent through the channel in a computationally unbounded way. Therefore not only do we have to guarantee that Alice and Bob each obtains a string that is close to private and uniform, but also we need to make sure that  they agree on this string with high probability. Following \cite{KR09, ChandranKOR10}, below we give the formal definition of a privacy amplification protocol in this case.

Let $w \in \bits^n$ be the secret string shard by Alice and Bob, and $w$ is sampled according to a distribution $W$. Let Protocol $(P_A, P_B)$ be executed in the presence of an active adversary Eve. Let $V_a$ denote the random variable that describes Alice's view of the communication when $(P_A, P_B)$ is executed and define $V_b$ likewise. We use small letters $v_a, v_b$ to denote specific values of $V_a, V_b$. The private randomness of Alice and Bob are denoted by $x$ and $y$ respectively. Alice's output is denoted by $r_A=P_A(w, v_a, x)$ and Bob's output is denoted by $r_B=P_B(w, v_b, y)$ (if successful, both outputs will be of length $\lambda_k$; rejection will be denoted by symbol $\perp$). Let $V$ denote Eve's view of the protocol. Since Eve is computationally unbounded, we can simply assume that Eve is deterministic.

\BD \cite{KR09, ChandranKOR10} An interactive protocol $(P_A, P_B)$ played by Alice and Bob on a communication channel fully controlled by an active adversary Eve, is a $(k, \lambda_k, \delta, \e)$-\emph{privacy amplification protocol} if it satisfies the following properties whenever $H_\infty(W) \geq k$:
\begin{enumerate}
\item \underline{Correctness.} If Eve is passive, then $\Pr[R_A=R_B]=1$.
\item \underline{Robustness.} The probability that the following experiment outputs ``Eve wins" is at most $\delta$: sample $w$ from $W$; let $v_a, v_b$ be the communication upon execution of $(P_A, P_B)$ with Eve actively controlling the channel, and let $r_A=P_A(w, v_a, x)$, $r_B=P_B(w, v_b, y)$. Output ``Eve wins" if $(r_A \neq r_B \land r_A \neq \perp \land r_B \neq \perp)$.
\item \underline{Extraction.} Letting $V$ denote Eve's view of the protocol, 

\[|(R_A, V | R_A \neq \perp)-(U_{\lambda_k}, V)| \leq \e\]

and

\[|(R_B, V | R_B \neq \perp)-(U_{\lambda_k}, V)| \leq \e.\]
\end{enumerate} 

$\ell = \log (1/\delta)$ is called the \emph{security parameter} of the protocol.
\ED     

To build such protocols, an important ingredient is an interactive authentication protocol. In such a protocol, Alice takes a message $m$ as input and tries to authenticate the message to Bob over the channel. Bob obtains message $m_B$ at the end of the protocol. We now give the formal definition of such a protocol.

\BD \cite{KR09, ChandranKOR10} An interactive protocol $(P_A, P_B)$ played by Alice and Bob on a communication channel fully controlled by an active adversary Eve, is a $(k, \ell)$-\emph{interactive authentication protocol} if it satisfies the following properties whenever $H_\infty(W) \geq k$:
\begin{enumerate}
\item \underline{Correctness.} If Eve is passive, then $\Pr[m_B=m]=1$.
\item \underline{Robustness.} The probability that the following experiment outputs ``Eve wins" is at most $2^{-\ell}$: sample $w$ from $W$; let $v_a, v_b$ be the communication upon execution of $(P_A, P_B)$ with Eve actively controlling the channel, and let $m_B=P_B(w, v_b, y)$. Output ``Eve wins" if $(m_B \neq m \land m_B \neq \perp)$.

\end{enumerate} 

Again $\ell$ is called the \emph{security parameter} of the protocol.

\ED

It is shown in \cite{KR09} that if we have a $(k, \ell)$ interactive authentication protocol for messages of length $\Theta(\ell+\log n)$, then it can be used to construct a $(k, \lambda_k, 2^{-\ell}, \e)$ privacy amplification protocol.  

As shown in the definitions, in the setting of privacy amplification with an active adversary, typical assumptions are that Alice and Bob have access to local (non-shared) uniform random bits. The first protocol given for this problem is due to Maurer and Wolf \cite{MW97}, which works as long as $W$ has entropy $k > \frac{2n}{3}$. This was later improved to $k > \frac{n}{2}$ in \cite{DKRS06}. Both of these protocols are single-rounded. In \cite{DW09} it is shown that no single round protocol exists if the secret $W$ has entropy $k < \frac{n}{2}$.

The first protocol that can handle $k < n/2$ (in fact, any $k > \polylog(n)$) is due to Renner and Wolf \cite{RW03}. In that paper they constructed a protocol using $\Theta(\ell+\log n)$ rounds of interaction, where $\ell$ is the security parameter. Later, several improvements appeared \cite{KR09, DW09, ChandranKOR10}. These papers intend to optimize different parameters in the protocol. For example, although achieving the same asymptotical behavior, \cite{KR09} considerably improves the constants hidden in $\Theta$, and thus makes the protocol much more practical. \cite{DW09} reduces the number of rounds needed to just two. \cite{ChandranKOR10} improves the entropy loss of the protocol to $\Theta(\ell+\log n)$, which is asymptotically optimal.  

In this paper, we seek to minimize the requirements on the local randomness used by Alice and Bob. Specifically, we ask the following two questions.

\textbf{Question 1: What is the minimum number of  random bits that Alice and Bob have to use, to ensure a protocol of security parameter $\ell$?} 

For this question, previous results all require at least $\Theta((\ell+\log n) \log n)$ local random bits for Alice and Bob. Specifically, the works of \cite{RW03}, \cite{KR09} and \cite{DW09} require $\Theta((\ell+\log n)^2)$ bits. The work of \cite{ChandranKOR10} requires $\Theta((\ell+\log n) \log n)$ bits. We note that non-constructively, by using the non-malleable extractor proposed in \cite{DW09}, there is a protocol that simultaneously achieves the optimal randomness complexity of $\Theta(\ell +\log n)$, the optimal round complexity of two and the optimal entropy loss of $\Theta(\ell +\log n)$.

\textbf{Question 2: Can Alice and Bob still agree on a \emph{nearly uniform} secret key, if they only have access to local weak random sources, instead of truly uniform random bits?}

For this question, as far as we know, there are no results that give a positive answer until now. All previous results require that Alice and Bob have access to local truly uniform random bits.

These questions are natural and important, for the same reasons that we have discussed above. First, it is not clear how to obtain uniform random bits in the real world. Not to mention a large number of uniform random bits. Thus we would like to know what is the minimum number of  random bits that Alice and Bob need. Second, it is most likely that Alice and Bob only have access to local weak random sources instead of truly random bits, either because weak random sources are the best they can get, or because their local random bits are compromised by side channel attacks. Thus we would like to know whether privacy amplification is still possible in this case. We feel that this is a new and important direction, as all previous works are taking for granted that Alice and Bob have access to a large number of truly uniform random bits, which could be far from true in practice. In this paper, we make the first effort towards minimizing the requirements on the local randomness used by Alice and Bob.


\subsection{Our Contribution}

For the first question, we show that $\Theta(\ell+\log n)$ local random bits suffice to achieve security parameter $\ell$, as long as the entropy of $W$ is at least $n^{\beta}$ for an arbitrary constant $0<\beta<1$. Thus our result improves the previous best result by a $\log n$ factor. This is also asymptotically optimal, because $\Omega(\ell+\log n)$ bits are needed to extract random bits from a general weak random source and achieve error $2^{-\ell}$. Specifically, we have the following theorem.

\BT \label{thm:main1} For all positive integers $n, \ell$ and every $0<\gamma<\beta<1$, assume that Alice and Bob share an $(n,k)$ weak random source $W$ with $k \geq n^{\beta}$. Then there exists an efficient $(k, \lambda_k, 2^{-\ell}, \e)$ privacy amplification protocol, where the total number of random bits that Alice and Bob use is $\Theta(\ell+\log n)$. The entropy loss of the protocol is $n^{\gamma}(\ell+\log n)$ and the length of the extracted key is $\lambda_k=k-n^{\gamma}(\ell+\log n)-O(\log (1/\e))$.

\ET

\begin{remark}
Note that in the case where $\ell > n^{\gamma}$, our protocol actually achieves better entropy loss than those of \cite{RW03, KR09, DW09}.

\end{remark}

For the second question, we show that 

\begin{enumerate}
\item Non constructively, we can do as good as if Alice and Bob have access to local random bits. Specifically, if Alice and Bob have two independent $(n,k)$ sources and they share an independent $(n,k)$ source, then there is a (possibly inefficient) protocol that achieves privacy amplification up to security parameter $\Omega(k)$.

\item If Alice and Bob have two independent $(n,(\frac{1}{2}+\delta)n)$ sources and they share an independent $(n,k)$ source, then there is an explicit protocol that achieves privacy amplification up to security parameter $k^{\Omega(1)}$.

\item If Alice and Bob have two independent $(n,\delta n)$ sources and they share an independent $(n,k)$ source, then there is an explicit protocol that achieves privacy amplification up to security parameter $\Omega(\log k)$.

\end{enumerate}

Specifically, we have the following theorems.

\BT \label{thm:main2} For all positive integers $n, k$ where $k > \log(n)$, assume that Alice and Bob have two independent local $(n, k)$ sources, and they share an independent $(n,k)$ source $W$. Then non-constructively there exists a $(k, k-O(\log n+\log (1/\delta)+\log (1/\e)), \delta, \e)$ privacy amplification protocol.

\ET

\BT \label{thm:main3} For all positive integers $n, k$ where $k \geq \polylog(n)$ and any constant $0<\delta<1$, assume that Alice and Bob have two independent local $(n, (1/2+\delta)n)$ sources, and they share an independent $(n,k)$ source $W$. Then there exists an efficient $(k, k-k^{\Omega(1)}, 2^{-k^{\Omega(1)}}, 2^{-k^{\Omega(1)}})$ privacy amplification protocol.

\ET

\BT \label{thm:main4} For all positive integers $n, k$ where $k \geq \polylog(n)$ and any constant $0<\delta<1$, assume that Alice and Bob have two independent local $(n, \delta n)$ sources, and they share an independent $(n,k)$ source $W$. Then there exists an efficient $(k, k-k^{\Omega(1)}, 1/\poly(k), 1/\poly(k))$ privacy amplification protocol.

\ET

Our results are the first results to give protocols that achieve privacy amplification when Alice and Bob only have access to local weak random sources.

\tableref{table:result} summarizes our results compared to some previous results, assuming the security parameter is $\ell$.

\begin{table}[ht] 
\centering 
\begin{tabular}{|l|l|l|l|l|} 
\hline Construction &  Entropy of $W$  &  Local randomness & Rounds & Entropy loss \\ 
\hline Optimal, non-explicit & $k > \log n$ & $\Theta(\ell + \log n)$ bits & $2$ & $\Theta(\ell + \log n)$ \\
\hline Optimal, non-explicit & $k > \log n$ & $(n, k > \log n)$-source & $2$ & $\Theta(\ell + \log n)$ \\
\hline \cite{MW97} & $k > 2n/3$ & $(n-k)$ bits & $1$ & $(n-k)$ \\
\hline \cite{DKRS06} & $k > n/2$ & $(n-k)$ bits & $1$ & $(n-k)$ \\
\hline \cite{RW03, KR09} & $k \geq \polylog(n)$ & $\Theta((\ell + \log n)^2)$ bits & $\Theta(\ell + \log n)$ & $\Theta((\ell + \log n)^2)$ \\ 
\hline \cite{DW09} & $k \geq \polylog(n)$ & $\Theta((\ell + \log n)^2)$ bits & $2$ & $\Theta((\ell + \log n)^2)$ \\ 
\hline \cite{ChandranKOR10} & $k \geq \polylog(n)$ & $\Theta((\ell + \log n)\log n)$ bits & $\Theta(\ell + \log n)$ & $\Theta(\ell + \log n)$ \\ 
\hline This paper & $k \geq n^{\beta}$ & $\Theta(\ell + \log n)$ bits & $\Theta(\ell + \log n)$ & $n^{\gamma}(\ell+\log n)$\\  
\hline This paper & $k \geq \polylog(n)$ & $(n, (1/2+\delta)n))$-source & $2$ & $k^{\Omega(1)}, \ell = k^{\Omega(1)}$ \\
\hline This paper & $k \geq \polylog(n)$ & $(n, \delta n)$-source & $\Theta(\log k)$ & $k^{\Omega(1)}, \ell = \Omega(\log k)$ \\ \hline
\end{tabular}
\caption{\textbf{Summary of Results on Privacy Amplification with an Active Adversary}} 
\label{table:result}
\end{table}

\subsection{Overview of the Constructions and Techniques}
Here we give a brief discussion of the constructions and the techniques that we use. First we show how we reduce the number of random bits used. For simplicity we focus on the case where $\ell=\Omega(\log n)$. Thus to use the protocols in \cite{KR09}, Alice needs to authenticate $\Theta(\ell)$ bits to Bob and we want to use only $\Theta(\ell)$ local random bits. 

\subsubsection{Reducing the Number of Local Random Bits}
Our starting point is the protocol in \cite{ChandranKOR10}, which builds upon the protocols in \cite{RW03, KR09}. Let's first briefly review that protocol. First by the results in \cite{RW03, KR09}, to achieve security parameter $\ell$ it suffices to give an interactive authentication protocol in which Alice an authenticate $\Theta(\ell+\log n)$ bits to Bob, such that the probability that Eve can successfully change the message is at most $2^{-\ell}$. To do this, the protocol in \cite{ChandranKOR10} starts by encoding the message $m$ into an error correcting code for edit errors. The codeword has length $O(\ell)$ and the edit distance between any two different codewords is $\Omega(\ell)$.

Next, in the protocol Alice sends the encoded message to Bob bit by bit. Sending each bit takes one round. In each round, Alice and Bob each sends out a fresh random seed. The seed is used in a strong extractor to extract some constant $c$ number of random bits from $W$. The analysis goes by arguing that for Eve to change the codeword to a different one, Eve has to make $\Omega(\ell)$ edit operations. Among these operations a constant fraction will result in Eve answering a challenge (coming up with a fresh output of the extractor), which Eve can only succeed with probability roughly $2^{-c}$. Therefore the overall success probability of Eve is roughly $(2^{-c})^{\Omega(\ell)}=2^{-\Omega(\ell)}$. Since the seed for an extractor has to be at least $\log n$ bits long, the total random bits used by Alice or Bob is $\Omega((\ell+\log n) \log n)$. 

Looking carefully into the analysis, we see that this protocol is actually wasting random bits in the following sense: in each round it uses $\log n$ fresh random bits, but if Eve has to answer a challenge, her success probability only goes down by a factor of $2^{-c}$ for some constant $c>1$. On the other hand, we know that $\Omega(\log n)$ bits suffice to extract random bits from $W$ and achieve error $1/\poly(n)$. Thus intuitively we would like Eve's success probability to go down by a factor of $1/\poly(n)$ each time she answers a challenge (note that the factor cannot be smaller than the error of the extractor). Another way of saying this is that if somehow magically we have extractors with seed length $O(1)$ that achieve error $2^{-c}$, then the number of random bits that Alice and Bob need would be $O(\ell)$, which is optimal. However of course it's impossible to build such extractors.

Thus we want to fully exploit the random bits used. Suppose we can use $\Theta(t)$ bits ($t=\Omega(\log n)$) to make Eve's success probability go down by a factor of $2^{-\Omega(t)}$ (this is the best we can hope for, since this is the error of the extractor with seed length $O(t)$) each time she answers a challenge, then we only need $\Theta(\ell/t)$ steps to make it decrease to $2^{-\ell}$. Thus the random bits needed will be $\Theta(t) \cdot \Theta(\ell/t)=\Theta(\ell)$.  

To achieve this goal, we design a new way for Alice to authenticate a message to Bob bit by bit. Again we encode the message $m$ into an error correcting code for edit errors, as in \cite{ChandranKOR10}. The codeword has length $\Theta(\ell)$. Now our protocol proceeds in phases. In each phase Alice and Bob use $\Theta(t)$ fresh random bits, for some chosen parameter $t =\Omega(\log n)$. Since we want the total number of random bits used to be $\Theta(\ell)$, we can only use $\Theta(\ell/t)$ phases. Thus in each phase Alice needs to send $\Theta(t)$ bits to Bob. We actually design the protocol such that in each phase, Alice sends exactly $t$ bits to Bob.

In each phase, the protocol does the following. In the beginning Alice sends $t$ fresh random bits $X$ to Bob. Bob also sends $t$ fresh random bits $Y$ to Alice. Of course what they actually receive may be different, say $Y'$ for Alice and $X'$ for Bob. Alice then computes $\Ext(W, Y')$ and Bob computes $\Ext(W, X')$, where $\Ext$ is a strong seeded extractor. Locally Alice computes the correct $\Ext(W, X)$ and Bob computes the correct $\Ext(W, Y)$. 

We then define three set of integers $C_{1i}=2^{3i-2}t, C_{2i}=2^{3i-1}t, C_{3i}=2^{3i}t$ for $i=1,\cdots, t$. Note that a phase consists of $t$ rounds and in round $i$ we are going to use the three numbers $C_{1i}, C_{2i}, C_{3i}$ to transfer a bit from Alice to Bob. Here's how we do it: if the bit is 0, Alice sends 0 and the prefix of $\Ext(W, Y')$ of length $C_{1i}$ to Bob; if the bit is 1 then Alice sends 1 and the prefix of $\Ext(W, Y')$ of length $C_{2i}$ to Bob. Bob then receives the message, and checks if the prefix matches the prefix of $\Ext(W, Y)$ in either case. If it doesn't match, Bob will abort and output $\perp$. Otherwise Bob responds to Alice with the prefix of $\Ext(W, X')$ of length $C_{3i}$ and Alice checks if the prefix matches the prefix of $\Ext(W, X)$. If it doesn't match, Alice will abort and output $\perp$. If neither of them aborts then this round is over and they move on to the next round.

We now want to argue that some of Eve's edit operations will result in Eve answering a challenge, and Eve's success probability of answering a challenge is roughly $2^{-\Omega(t)}$. Intuitively, the reason is as follows. Note that the numbers $C_{1i}, C_{2i}, C_{3i}$ have the property that if we arrange them into a sequence $C_{11}, C_{21}, C_{31}, C_{12}, C_{22}, C_{32}, \cdots$, then the next number is always twice as large as the previous number. This means that the length of the prefix in the protocol increases by a factor of 2 each time. Now take for example that Eve is making an insert operation in round $i$, then Bob is expecting the prefix of $\Ext(W, Y)$ of length at least $C_{1i}$ (in the 0 case). On the other hand, the total number of bits of information that Eve can use is at most $C_{3(i-1)}+C_{2(i-1)}=3C_{1i}/4$ (a prefix of $\Ext(W, X')$ and a prefix of $\Ext(W, Y')$. Note that $X$ and $Y$ here are independent of $\Ext(W, Y)$ by the strong extractor property, thus they do not give any information of $\Ext(W, Y)$. So are the bits that Alice tries to send to Bob). Since $\Ext(W, Y)$ is close to uniform Eve has to come up with at least $C_{1i}-3C_{1i}/4=C_{1i}/4 \geq t/2$ new random bits. Therefore Eve's success probability is roughly $2^{-\Omega(t)}$. The case of deletion and changing 0 to 1 are similar, because we have chosen $\{C_{1i}\}, \{C_{2i}\}, \{C_{3i}\}$ such that a number is always greater than the sum of its previous two numbers. The only case where Eve might succeed with probability 1 is the case of changing from 1 to 0. However as in the context of \cite{KR09, ChandranKOR10} the number of 1's in the message is known to Bob. Therefore Eve cannot always make operations of changing from 1 to 0.

The actual analysis is more complicated, since after some operations made by Eve, the round number of Alice and Bob may not be the same. In fact they might not even be in the same phase and we have to do the analysis in Eve's view. Nevertheless the above discussion captures the main ideas behind the analysis. In Eve's view, we define a phase as the rounds from the one where either Alice or Bob starts a new phase, to the one before the next round where either Alice or Bob starts a new phase. The key point here is that in a new phase, either Alice or Bob will announce a fresh random seed. Suppose Alice does and the fresh random seed is $X$. Now conditioned on the fixings of all previous transcript, $X$ is uniform, $W$ has a lot of entropy left and is independent of $X$. Thus $\Ext(W,X)$ is close to uniform conditioned on all previous transcript. Hence whenever Eve has to answer a challenge, her success probability is at most $2^{-\Omega(t)}$ by the above analysis. Now again by the property of the error correcting code, Eve has to make $\Omega(\ell)$ edit operations to change one codeword to another. Intuitively these operations will go into $\Omega(\ell/t)$ phases. We again show that for a constant fraction of these phases, Eve has to answer at least one challenge. Thus the overall success probability of Eve is at most $(2^{-\Omega(t)})^{\ell/t}=2^{-\Omega(\ell)}$.

Note that the total entropy loss of the protocol is something like $2^{3t} \ell$. Thus we can choose $t=\Theta(\log n)$ and the entropy loss of the protocol will be $n^{\gamma} \ell$ for an arbitrary constant $0<\gamma<1$. Therefore we need the entropy of $W$ to be at least $n^{\beta}$ for an arbitrary constant $0<\beta<1$.

\subsubsection{Using Local Weak Random Sources}
Here what we do is to try to reduce the case to where Alice and Bob have access to local private random bits. In other words, we want to design a protocol such that at the end of the protocol, Alice and Bob end up with nearly private and random bits, while their shared secret $W$ still has a lot of entropy left. Non-constructively, this is simple, because non-constructively there exists strong two-source extractors for min-entropy as small as $k >\log n$. If we have a strong two-source extractor, then Alice and Bob just each applies this extractor to his or her own source and $W$. By the property of the strong two-source extractor, even conditioned on $W$, their outputs are close to uniform. Moreover, conditioned on $W$ their outputs are deterministic functions of their own sources, and are thus independent. Eve also knows nothing about their outputs since all computations are private. Thus we are done. In another case where Alice and Bob each has an independent source with entropy $k = (1/2+\delta)n$, a construction of Raz \cite{Raz05} serves as a strong two source extractor. Thus in this case we have an explicit protocol.

The hard case is where Alice and Bob only have independent sources with entropy $k = \delta n$. Our starting point here is how we can construct an extractor for these three sources $X, Y$ and $W$. In other words, let's first forget about the communication problem and see how we can get a 3-source extractor. 

Since $X$ has linear min-entropy, a standard approach would be to convert $X$ into a somewhere high entropy (say entropy rate $0.9$) source $\bar{X}$, using the condenser based on the sum-product theorem \cite{BarakKSSW05, Zuc07}. $\bar{X}$ is a matrix with a constant number of rows such that at least one of the rows has entropy rate $0.9$. Once we have this, we can apply Raz's extractor to each row of $\bar{X}$ and $W$, and we get a somewhere random source with a constant number of rows. Now we can extract from such a source and an independent weak random source using the two-source extractor in \cite{BarakRSW06}.

So now how do we use these ideas in the case where Alice and Bob are separated by a channel controlled by an active adversary Eve? As a first step, we still convert $X$ and $Y$ into somewhere high entropy (say entropy rate $0.9$) sources $\bar{X}$ and $\bar{Y}$ with $D=O(1)$ rows. Next we apply Raz's extractor to each row of $\bar{X}$ and $W$, and each row of $\bar{Y}$ and $W$. Thus we get two somewhere random sources $SR_x$ and $SR_y$. Note that since Raz's extractor is a strong two-source extractor, the random rows in $SR_x$ and $SR_y$ are close to being independent of $W$. This is important for us.

Next, we have Alice authenticate a string to Bob. To do this, we want to use the authentication protocol we discussed in the previous section. However, now Alice and Bob don't have access to random bits. The important observation here is that they have \emph{somewhere random sources}. In particular, the random rows of $SR_x$ and $SR_y$ can be used as seeds for a strong seeded extractor in the authentication protocol (since they are independent of $W$). Of course we don't know which row is the random row, thus we take a slice from the somewhere random sources with small width (so that $X$ and $Y$ don't lose much entropy) and use these slices in the authentication protocol. We call these slices $X_1$ and $Y_1$. How do we use them? As before Alice and Bob announce their slices to each other and each time they communicate, they compute prefixes of the outputs of an extractor. They then check if the prefixes they received match the prefixes they compute locally. Only this time we apply the extractor to $W$ using each row of the slice as a seed. Thus the output of the extractor is also a matrix of $D$ rows. Since the random row of the slice is close to being independent of $W$, the output of the extractor is also somewhere random. Next, instead of increasing the length of the prefix by a factor of 2 each time, we increase the length of the prefix by a factor of $2D$, because each time Alice or Bob reveals a matrix of $D$ rows. Now as before to answer a challenge Eve has to come up with the random row in the output of the extractor, whose length is larger than the total number of bits revealed so far. Therefore Eve can only succeed with a small probability. One thing to note here is that we don't use the error correcting code as in the previous section. All we need is to make sure that Eve has to answer at least one challenge. Given this protocol, Alice uses it to send another small slice of $SR_x$ to Bob, and Bob uses this slice to extract random bits from his own source. We call this slice $X_2$.

There are two  problems with the above discussion. First, the small slice $X_2$ sent by Alice may not be independent of $W$ or the random row of the extractor output. Second, since each time the length of the prefix increases by a factor of $2D$ and each time we want fresh entropy in the extractor output, Alice can only send $\alpha \log k$ bits with some $\alpha <1$, where $k$ is the entropy of $W$. With this small number of bits it's not clear how Bob can extract random bits from his own source.

For the first problem, we show that although the small slice $X_2$ sent by Alice may not be independent of $W$ or the random row of the extractor output, with high probability over the fixings of $X_2$, the random row of the extractor output has very high min-entropy. This is mainly because the length of $X_2$ is very small compared to the extractor output. Thus a typical fixing of it doesn't reduce the entropy of the random row of the extractor output by much. Now we can show that with correct parameters, the high min-entropy row of the extractor output still suffices for authentication. Thus for a typical value of the small slice, the success probability of Eve changing it is still small. Note now Eve may be able to actually change a small probability mass of the slice sent by Alice, but that doesn't hurt us much. This is different from the case where Alice and Bob have local random bits. For the second problem, luckily Bob also has a somewhere random source $SR_y$. Thus we can take a small slice $Y_2$ of $SR_y$ so that the two-source extractor from \cite{BarakRSW06} can be used to extract random bits from these two sources. 

Now suppose that Bob correctly received the small slice $X_2$ sent by Alice, and Bob takes a small slice $Y_2$ of his somewhere random source. Let the output of the \cite{BarakRSW06} extractor be $R$. We first fix $W$ and now $X_2$ and $Y_2$ are deterministic functions of $X$ and $Y$ respectively, and are thus independent. Moreover $X_2$ is somewhere random. Thus $R$ is close to uniform. Furthermore, since the two source extractor from \cite{BarakRSW06} is strong, we can now fix $Y_2$ and conditioned on this fixing, $R$ is still close to uniform. Now $R$ is a deterministic function of $X$. Note that now all strings revealed by Bob are functions of $Y_1$ and $X_1$ (since $W$ is fixed), and $Y_1$ is a deterministic function of $Y$ and has small size. Thus we can further fix $Y_1$ and conditioned on this fixing, $R$ is still close to uniform and is independent of $Y$. Moreover $Y$ has a lot of entropy left and all the strings revealed by Bob are now deterministic functions of $X_1$. Therefore now we can apply a strong seeded extractor to $Y$ and $R$ and Bob obtains $S_y$. Note that we can condition on $R$ and $S_y$ is still close to uniform by the strong extractor property. Now $S_y$ is a deterministic function of $Y$ and is thus independent of all the transcripts revealed so far, and $X$. Thus Bob has obtained random bits that are close to uniform and private.

We actually cheated a little bit above, because again the size of $R$ is very small compared to $Y$. Thus we won't be able to apply a seeded extractor to $Y$ and $R$. However we can fix this problem by taking a slice $Y_3$ of $SR_y$. The size of $Y_3$ is much larger compared to the length of the transcript, but much smaller compared to $Y$. It is actually a slice with width $k^{\Omega(1)}$ ($R$ will have size $\Omega(\log k)$). Since in the analysis we fix $W$, $Y_3$ is a deterministic function of $Y$, and the random row of $Y_3$ still has a lot of entropy left conditioned on the fixings of the transcript. Therefore we can apply the strong seeded extractor to $Y_3$ and $R$, and the above analysis about Bob obtaining private random bits still holds.

By symmetry Alice can also take a slice $X_3$ of $SR_x$ and apply a strong seeded extractor to $X_3$ and $R$, and the above argument would also work for Alice. Therefore now Bob can use the authentication protocol to send $R$ to Alice, and Alice applies the extractor to $X_3$ and $R$. By the same discussion above Eve may be able to change only a small probability mass of $R$, and this doesn't hurt us much. Thus at the end of the protocol Alice and Bob end up with nearly private and uniform random bits, while their shared secret $W$ still has a lot of entropy left. Thus we have reduced the problem to the case where Alice and Bob have access to local uniform random bits, and previous results can be used to construct a privacy amplification protocol. However since we only manage to send $\Omega(\log k)$ bits from Alice to Bob, the error of the extractor and thus the security parameter of the protocol is $\Omega(\log k)$.

\section{Roadmap} The rest of the paper is organized as follows. In \sectionref{sec:prelim} we give the preliminaries and previous works that we use. In \sectionref{sec:codes} we briefly describe the error correcting code that we use for edit errors, as in \cite{ChandranKOR10}. \sectionref{sec:randombits} gives the formal description of our protocol to achieve the optimal number of local random bits, and the analysis of the protocol. \sectionref{sec:weaksource} gives our protocols for privacy amplification with local weak random sources, and the analysis of these protocols. We conclude in \sectionref{sec:con} with some open problems. Finally in \appendixref{sec:thmsrg} we outline a proof sketch of \theoremref{thm:srgeneral}, which has slight differences in parameters from the theorem in \cite{BarakRSW06}.

\section{Preliminaries} \label{sec:prelim}
We use common notations such as $\circ$ for concatenation and $[n]$ for $\{1, 2, \cdots, n\}$. All logarithms are to the base 2. We often use capital letters for random variables and corresponding small letters for their instantiations.

\subsection{Basic Definitions}
\begin{definition} [statistical distance]Let $D$ and $F$ be two distributions on
a set $S$. Their {\bf statistical distance} is
\begin{align*}
|D-F|\eqdef \max_{T \subseteq S}(|D(T) - F(T)|) = \frac{1}{2}
\sum_{s \in S}|D(s)-F(s)|
\end{align*}
If $|D-F| \leq \e$ we say that $D$ is \emph{$\e$-close} to $F$.
\end{definition}

\begin{definition}
The \emph{min-entropy} of a random variable $X$ is defined as
\[H_{\infty}(X)=min_{x \in \supp(X)}\{-\log_2 \Pr[X=x]\}.\] We say
$X$ is an $(n,k)$-source if $X$ is a random variable on $\bits^n$
and $H_{\infty}(X) \geq k$. When $n$ is understood from the context we simply say that $X$ is a $k$-source.
\end{definition}

\subsection{Somewhere Random Sources, Extractors and Condensers}

\begin{definition} [Somewhere Random sources] \label{def:SR} A source $X=(X_1, \cdots, X_t)$ is $(r, t)$
  \emph{somewhere-random} (SR-source for short) if each $X_i$ takes values in $\bits^r$ and there is an $i$ such that $X_i$ is uniformly distributed.
\end{definition}

\BD
An elementary somewhere-k-source is a 	vector of sources $(X_1, \cdots, X_t)$, such that some $X_i$ is a $k$-source. A somewhere $k$-source is a convex combination of elementary somewhere-k-sources.
\ED

\BD
A function $C: \bits^n \times \bits^d \to \bits^m$ is a $(k \to l, \e)$-condenser if for every $k$-source $X$, $C(X, U_d)$ is $\e$-close to some $l$-source. When convenient, we call $C$ a rate-$(k/n \to l/m, \e)$-condenser.   
\ED

\BD
A function $C: \bits^n \times \bits^d \to \bits^m$ is a $(k \to l, \e)$-somewhere-condenser if for every $k$-source $X$, the vector $(C(X, y)_{y \in \bits^d})$ is $\e$-close to a somewhere-$l$-source. When convenient, we call $C$ a rate-$(k/n \to l/m, \e)$-somewhere-condenser.   
\ED

\begin{definition}\label{def:strongext}

A function $\Ext : \bits^n \times \bits^d \rightarrow \bits^m$ is  a \emph{strong seeded extractor} for min-entropy $k$ and error $\e$ if for every min-entropy $k$ source $X$, 

\[ |(\Ext(X, R), R)-(U_m, R)| < \e,\]
where $R$ is the uniform distribution on $d$ bits independent of $X$, and $U_m$ is the uniform distribution on $m$ bits independent of $R$. 
\end{definition}

\begin{definition}

A function $\TExt : \bits^{n_1} \times \bits^{n_2} \rightarrow \bits^m$ is  a \emph{strong two source extractor} for min-entropy $k_1, k_2$ and error $\e$ if for every independent  $(n_1, k_1)$ source $X$ and $(n_2, k_2)$ source $Y$, 

\[ |(\TExt(X, Y), X)-(U_m, X)| < \e\]

and

\[ |(\TExt(X, Y), Y)-(U_m, Y)| < \e,\]
where $U_m$ is the uniform distribution on $m$ bits independent of $(X, Y)$. 
\end{definition}

\subsection{Previous Work that We Use}

We are going to use condensers recently constructed based on the sum-product theorem. The following construction is due to Zuckerman \cite{Zuc07}.

\BT [\cite{Zuc07}] \label{thm:swcondenser}
For any constant $\beta, \delta>0$, there is an efficient family of rate-$(\delta \to 1-\beta, \e=2^{-\Omega(n)})$-somewhere condensers $\zuc: \bits^n \to (\bits^m)^D$ where $D=O(1)$ and $m=\Omega(n)$. 

\ET

The following theorem is adapted from \cite{BarakRSW06}.

\begin{theorem} [General Source vs Somewhere random source with few rows Extractor \cite{BarakRSW06}] \label{thm:srgeneral}
For every $n,k(n)$ with $k >
\log^{10}n,$ and constant $D>1$, there is a polynomial
time computable function $\SRGExt:\{0,1\}^{n} \times
\{0,1\}^{Dk} \rightarrow \{0,1\}^m $ s.t. if $X$ is an
$(n,k)$ source and $Y$ is a $(D \times k)$
-SR-source,

\[ | (Y , \SRGExt(X,Y)) - (Y , U_m) | < \epsilon \]
and
\[ | (X , \SRGExt(X,Y)) - (X , U_m) | < \epsilon, \]
where $U_m$ is independent of $X,Y$, $m = \Omega(k)$ and
$\epsilon = 2^{-\Omega(k)}$.
\end{theorem}

Note that the parameters here are slightly different from the statement in \cite{BarakRSW06}. For a proof sketch see the \appendixref{sec:thmsrg}.

\begin{theorem}[\cite{Raz05}] \label{thm:razweakseed} For any
  $n_1,n_2,k_1,k_2,m$ and any $0 < \delta < 1/2$ with

\begin{itemize}
\item $n_1 \geq 6 \log n_1 + 2 \log n_2$ \item $k_1 \geq (0.5 +
\delta)n_1 + 3 \log n_1 + \log n_2$ \item $k_2 \geq 5 \log(n_1 -
k_1)$ \item $m \leq \delta \min[n_1/8,k_2/40] - 1$
\end{itemize}

There is a polynomial time computable strong 2-source extractor
$\Raz : \bits^{n_1} \times \bits^{n_2} \rightarrow \bits^m$ for
min-entropy $k_1, k_2$ with error $2^{-1.5 m}$.

\end{theorem}

For a strong seeded extractor with optimal parameters, we use the following extractor constructed in \cite{GuruswamiUV07}.

\BT [\cite{GuruswamiUV07}] \label{thm:optext} 
For every constant $\alpha>0$, and all positive integers $n,k$ and $\e>\exp(-n/2^{O(\log^* n)})$, there is an explicit construction of a strong $(k,\e)$ extractor $\Ext: \bits^n \times \bits^d \to \bits^m$ with $d=O(\log n +\log (1/\e))$ and $m \geq (1-\alpha) k$.
\ET

We need the following simple lemma about statistical distance.

\BL \label{lem:convert} Assume we have 3 random variables $X_1, Y_1, Y_2$ such that $|Y_1-Y_2| \leq \e$. Then there exists a random variable $X_2$ with the same support of $X_1$, such that

\[|(X_1, Y_1)-(X_2, Y_2) | \leq \e.\]

\EL

\begin{proof}
We construct the random variable $X_2$ and the distribution $(X_2, Y_2)$ as follows. For any $y$, consider $\Pr[Y_1=y]$, $\Pr[Y_2=y]$ and the distribution $(X_1, Y_1=y)$. Let $\delta = \Pr[Y_1=y]-\Pr[Y_2=y]$. If $\delta \geq 0$,  then we do the following:

\begin{enumerate}
\item Define an arbitrary order on the set of the support of $X_1$.
\item While $\delta > 0$, pick a new $x$ from the support according to the above order and let $p=\Pr[X_1=x, Y_1=y]$.
\item Let $\Pr[X_2=x, Y_2=y]=p-min(p, \delta)$.
\item Let $\delta=\delta-min(p, \delta).$
\item When $\delta=0$, for all the rest $x$, let $\Pr[X_2=x, Y_2=y]=\Pr[X_1=x, Y_1=y]$.

\end{enumerate}

If $\delta <0$, then we do the following:
\begin{enumerate}
\item Pick an arbitrary $x$ from the support of $X_1$ and let $p=\Pr[X_1=x, Y_1=y]$.
\item Let $\Pr[X_2=x, Y_2=y]=p-\delta$.
\item For all the other $x$ in the support of $X_1$, let $\Pr[X_2=x, Y_2=y]=\Pr[X_1=x, Y_1=y]$.

\end{enumerate}

It is easy to see that the distribution $(X_2, Y_2)$ has marginal distribution $Y_2$ and $|(X_1, Y_1)-(X_2, Y_2) |=|Y_1-Y_2| \leq \e$.

\end{proof}

We are going to use the following standard lemma about conditional min-entropy.

\begin{lemma}[\cite{MW97}] \label{lem:condition} 
Let $X$ and $Y$ be random variables and let ${\cal Y}$ denote the
range of $Y$. Then for all $\e>0$
\[\Pr_Y \left [ H_{\infty}(X|Y=y) \geq H_{\infty}(X)-\log|{\cal Y}|-\log \left( \frac{1}{\e} \right )\right ] \geq 1-\e\]
\end{lemma}

In \cite{KR09}, it is shown an interactive authentication protocol can be used to construct a privacy amplification protocol. Specifically, we have the following theorem.

\BT [\cite{KR09}] \label{thm:authprivacy} Suppose there exists an efficient $(k, \ell)$ interactive authentication protocol for messages of length $\Theta(\ell+\log n)$, then there exits an efficient $(k, \lambda_k, 2^{-\ell}, \e)$ privacy amplification protocol.  

\ET

\section{Edit Distance Codes} \label{sec:codes}
We are going to use a family of codes designed for edit distance errors, as in \cite{ChandranKOR10}. We have the following definitions.

\BD \cite{ChandranKOR10} For any two strings $c$ and $c'$ of length $\lambda_c$, let $\EditDis(c, c')$ denote the edit distance between $c$ and $c'$, i.e., the number of single-bit insert and delete operations required to change string $c$ into $c'$. 

\ED

\BD \cite{ChandranKOR10} Let $m \in \bits^{\lambda_m}$. For some constant $0< e <1$, a function $\Edit: \bits^{\lambda_m} \to \bits^{\lambda_c}$ is a $(\lambda_m, e, \rho)$-error-detecting code for edit errors, if $\rho \lambda_c = \lambda_m$ and the following properties are satisfied:
\begin{itemize} 
\item $c=\Edit(m)$ can be computed in polynomial (in $\lambda_m$) time, given $m$, for all $m \in \bits^{\lambda_m}$.
\item For any $m, m' \in \bits^{\lambda_m}$ with $m \neq m'$, $\EditDis(c, c') \geq e \lambda_c$, where $c=\Edit(m)$ and $c'=\Edit(m')$. 
\end{itemize}

$\rho =\frac{\lambda_m}{\lambda_c}$ is called the rate of the code.
\ED

As in \cite{ChandranKOR10} the code we use is due to Schulman and Zuckerman \cite{SchulmanZ99}:

\BT  [\cite{SchulmanZ99, ChandranKOR10}]\label{thm:editcode}
Let $0<e<1$ be a constant. Then for some constant $0<\rho<1$ there exists a $(\lambda_m, e, \rho)$-error-detecting code for edit errors. Moreover, each codeword has the same Hamming weight (the number of 1's).
\ET

\section{Reducing the Number of Local Radom Bits} \label{sec:randombits}
In this section we show how we can reduce the number of local random bits used. First we give a new authentication protocol, that can be used to authenticate a short message. In this protocol Alice and Bob only announce a fresh random string to each other at the beginning. We need the following definition.

\BD
Given a $n$ bit string $r=r_1 \cdots r_n$, define $\Pre(r,s)$ to be the prefix of $r$ of length $s$. Define $r_[a:b]$ to be the substring $r_a \cdots r_b$ of $r$. Let $\mathsf{wt}(r)$ stand for the weight of $r$, i.e., the number of $1$'s in $r$.

\ED

Now we have the following protocol.

$\mathsf{Protocol}$ $\SAuth(w,m,t)$:
\begin{itemize}
\item Alice and Bob share an $n$-bit secret random string $w$ with min-entropy $k$. 
\item Alice wishes to authenticate a $t$-bit string $m=m_1 \cdots m_t$ to Bob. 
\item Let $\Ext$ be a strong extractor as in \theoremref{thm:optext}, set up to use $O(t)$ bits to extract $2^{3i}t $ bits that are $2^{-\Omega(t)}$-close to uniform.
\item Let $c>1$ be some integer. Define three set of integers as $C_{1i}=2^{3i-2}t, C_{2i}=2^{3i-1}t, C_{3i}=2^{3i}t$, where $i=1, \cdots, t$. 
\end{itemize}

\begin{enumerate}

\item Alice sends Bob a fresh random seed $x$, and Bob sends Alice a fresh random seed $y$.

\item Alice receives $y$ and computes $r_y=\Ext(w,y)$; Bob receives $x$ and computes $r_x=\Ext(w,x)$.

\item For $i=1$ to $t$ do the following

\begin{description}
\item[Step a]: If $m_i=0$, Alice sends $(0, \Pre(r_y, C_{1i}))$ to Bob. Otherwise she sends $(1, \Pre(r_y, C_{2i}))$ to Bob.

\item[Step b]: Bob receives the message and verifies $\Pre(r_y, C_{1i})=\Pre(\Ext(w,y), C_{1i})$ in the $0$ case and $\Pre(r_y, C_{2i})=\Pre(\Ext(w,y), C_{2i})$ in the $1$ case. If the verification does not go through, abort. Bob then sends $\Pre(r_x, C_{3i})$ to Alice.

\item[Step c]: Alice receives the message and verifies $\Pre(r_x, C_{3i})=\Pre(\Ext(w,x), C_{3i})$.
\end{description}

\end{enumerate}

Now we want a protocol that can authenticate a long message, say a message with $\ell$ bits. We assume that the number of 1's in the message is known to Bob. We have the following protocol, where Alice and Bob execute protocol $\mathsf{SAuth}$ for $\ell/t$ times and each time Alice authenticates $t$ bits to Bob.

$\mathsf{Protocol}$ $\Auth(w,m,t)$:
\begin{itemize}
\item Alice and Bob share an $n$-bit secret random string $w$ with min-entropy $k$. 
\item Alice wishes to authenticate an $\ell$-bit string $m=m_1 \cdots m_{\ell}$ to Bob. 
\end{itemize}
\begin{enumerate}

\item For $j=1$ to $\ell/t$ do

Alice and Bob execute protocol $\mathsf{SAuth}(w, m_{[(j-1)t+1:jt]}, t)$.

\item When received $\ell$ bits, Bob verifies that the number of ones in the message is $\mathsf{wt}(m)$. Abort otherwise.

\end{enumerate}

Now we can describe our final protocol. 

$\mathsf{Protocol}$ $\NAuth(w,m)$:
\begin{itemize}
\item Alice and Bob share an $n$-bit secret random string $w$ with min-entropy $k$. 
\item Alice wishes to authenticate an $\ell$-bit string $m=m_1 \cdots m_{\ell}$ to Bob. 
\item Let $\Edit(m)$ stand for the edit distance encoding of the string $m$, as in \theoremref{thm:editcode}. 
\item Let $t>0$ be some integer parameter to be chosen later.

\end{itemize}

\begin{enumerate}
\item Alice sends Bob the message $m$. Let $m'$ denote the message received by Bob.

\item Alice and Bob execute protocol $\Auth(w,s,t)$ where $s=\Edit(m)$.

\item Let $s'$ stand for the string received by Bob. Bob computes $\Edit(m')$. If $s'=\Edit(m')$, Bob accepts $m'$ as the received message. Otherwise, Bob rejects and aborts.
\end{enumerate}

\subsection{Analysis of the protocol}

Let $D$ denote the edit distance between $\Edit(m)$ and $\Edit(m')$, where the operations are insertion and deletion, as in \theoremref{thm:editcode}. Let $D'$ denote the number of insertions, deletions and changing bits from $0$ to $1$ that Eve has to make to change $\Edit(m)$ to $\Edit(m')$. We have the following claim.


\BCM \label{clm:operation}
$D' \geq D/4.$
\ECM
\begin{proof}[Proof of the cliam]
Since the operations that Eve can do include insertion, deletion and changing bits, we first consider the edit distance under these operations. Let $\bar{D}$ denote the edit distance between $\Edit(m)$ and $\Edit(m')$, under insertion, deletion and changing bits (0 to 1 and 1 to 0). Since every operation of changing bits can be replaced by two operations of insertion and deletion (e.g., deleting a ``0" and inserting a ``1"), we have

\[\bar{D} \geq D/2.\]

Let $P_0$ stand for the operation of changing 0 to 1, let $P_1$ stand for the operation of changing 1 to 0, and let $P_2$ stand for the operation of insertion or deletion. Let $n_0$ denote the number of $P_0$ operations Eve made, let $n_1$ denote the number of $P_1$ operations Eve made, and let $n_2$ denote the number of $P_2$ operations Eve made. 

Now consider the sequence of operations that Eve made. It is a sequence of $P_0$, $P_1$ and $P_2$. Note that the number of $1$'s in $\Edit(m)$ and $\Edit(m')$ are the same. However if Eve makes $n_0$ $P_0$ operations and $n_1$ $P_1$ operations, then there will be $n_1-n_0$ more $0$'s in $\Edit(m')$. These $0$'s must be changed back to $1$'s by insertion and deletion. Since changing a bit must take one deletion and one insertion, we have

\[n_2 \geq 2(n_1-n_0).\]

Also by the definition of edit distance we have

\[n_0+n_1+n_2 \geq \bar{D}.\]

Thus we have

\[D'=n_2+n_0 \geq 3n_2/4+(n_1-n_0)/2+n_0 \geq (n_0+n_1+n_2)/2 \geq \bar{D}/2 \geq D/4.\]
\end{proof}

We now need the following two definitions.

\BD(Challenge) We say that Eve has to answer a \emph{challenge} if the following happens: Conditioned on the fixing of some random variable $V$, the information that Eve is left with (from the strings revealed by Alice and Bob) has $l$ bits, and now Eve has to come up with a (close to ) uniform random string with at least $l+2t$ bits to avoid detection.

\ED

\BD(Phases) In Alice or Bob's view, a \emph{phase} is the period of one execution of $\mathsf{Protocol}$ $\SAuth(w,m,t)$.  In Eve's view of the protocol, a \emph{phase} is defined as the rounds from the round where either Alice or Bob starts a new phase (announces a fresh random seed) to the round before the next round where  either Alice or Bob starts a new phase. In Eve's view, we say a phase is a \emph{bad phase} if it contains an operation of insertion, deletion or changing a bit from $0$ to $1$. We say a phase is a \emph{challenge phase} if in that phase Eve has to answer at lease one challenge. 

\ED

Now we have the following lemma.

\BL \label{lem:challenge}In any two adjacent bad phases, at least one of them is a challenge phase.
\EL

\begin{proof}
Here for the purpose of clear statements we shall use words like ``w.h.p." and ``close to uniform". We'll deal with probability errors in the proof of the main theorem.

Let's consider the first bad phase. It starts with a round where either Alice or Bob announces a fresh random seed.  Suppose Alice announces a fresh random seed $X$. This means that in Alice's view, she is starting a new phase. Now consider Bob. There are two cases. 

\textbf{Case 1:} Bob is also starting a new phase. Thus Bob also announces a fresh random seed $Y$. Now we can fix all previous seeds announced by Alice and Bob and $X$ and $Y$ are still uniform and independent. Now after this fixing all previous strings revealed by Alice and Bob are deterministic functions of $w$. Therefore we can further fix all these strings and conditioned on these fixings w.h.p. $W$ still has a lot of min-entropy left (we choose parameters that guarantee this happens). Thus $(\Ext(W, X), \Ext(W, Y)$ is close to uniform, even conditioned on $(X, Y)$ by \theoremref{thm:optext}. In this case we claim that the first operation of insertion, deletion or changing a bit from 0 to 1 will result in Eve's answering for a challenge.

To see this, note that before such an operation, the round numbers of this phase in Alice's and Bob's view are the same. Let $i$ be the round in which the first such operation takes place. Note that we have fixed all strings in previous phases.

\begin{itemize}
\item If the operation is an insertion, then Eve has to at least come up with $\Pre(\Ext(W, Y), C_{1i})$, which has $2^{3i-2}t$ bits. If $i=1$ then this is $2t$ bits and Eve has no other information about it. If $i>1$ then Eve has at most $C_{2(i-1)}+C_{3(i-1)}=3 C_{1i}/4$ bits. Thus 

\[C_{1i}-3 C_{1i}/4=2^{3i-2}t/4 > 2t.\]

\item If the operation is a deletion, then Eve has to at least come up with $\Pre(\Ext(W, X), C_{3i})$, which has $2^{3i}t$ bits. Meanwhile Eve has at most $C_{2i}+C_{3(i-1)}<3 C_{3i}/4$ bits of information. Thus 

\[C_{3i}-3 C_{3i}/4=2^{3i}t/4 \geq 2t.\]

\item If the operation is changing a bit from $0$ to $1$, then Eve has to come up with $\Pre(\Ext(W, Y), C_{2i})$, which has $2^{3i-1}t$ bits. If $i=1$ then Eve has at most $C_{1i}=C_{2i}/2$ bits of information. Thus

\[C_{2i}-C_{2i}/2=2^{3i-1}t/2 \geq 2t.\]

If $i>1$ then Eve has at most $C_{1i}+C_{3(i-1)}=3C_{2i}/4$ bits of information. Thus

\[C_{2i}-3C_{2i}/4=2^{3i-1}t/4 > 2t.\]

\end{itemize} 

Thus in this case the first bad phase is a challenge phase.

\textbf{Case 2:} Bob is in the middle of an old phase. In this case we claim that the only operation that Eve can make to avoid a challenge is the insertion operation.

To see this, consider all the strings that are revealed by Bob, and the strings that are going to be revealed by Bob until Bob enters a new phase. Except the random seeds that Bob have announced, these strings are all substrings of $\Ext(W, X')$, where $X'$ is a deterministic function of strings revealed before $X$ is revealed. In other words, Eve cannot change $X'$ while Bob is in the middle of a phase. Therefore $X$ is completely independent of all these strings. Thus we can fix all the strings (including the strings that Bob are going to reveal before he enters a new phase) and w.h.p. $W$ still has a lot of min-entropy left. Thus $\Ext(W, X)$ is close to uniform conditioned on all these fixings and the fixing of $X$.

Now if Eve is ever going to make a deletion, or changing a bit, then he has to come up with $\Pre(\Ext(W,X), C_{3i})$ as Bob's response, which has $C_{3i}$ bits. Meanwhile Eve has at most $C_{2i}=C_{3i}/2$ bits of information (since strings from Bob are fixed).  Thus

\[C_{3i}-C_{3i}/2=2^{3i}t/2 > 2t,\]
and this operation will be a challenge for Eve.

If so then again the first bad phase is a challenge phase. Otherwise the only thing Eve can do is to keep inserting until Bob enters a new phase. Now In Eve's view also a new phase begins, and it is a phase where Alice and Bob starts a new phase simultaneously (since inserting does not change the round number in Alice's view). Thus until the next bad phase Alice and Bob will always have the same round number during a phase, and the next bad phase will be a phase of Case 1. Therefore in this case the next bad phase will be a challenge phase.

The case where a new phase in Eve's view starts by Bob entering a new phase is similar and symmetric, we thus omit the details here.
\end{proof}
 
Now suppose in the protocol the code $\Edit(m)$ has length $\ell$. By the property of the code there exists a constant $\alpha>0$ s.t. the edit distance between $\Edit(m)$ and $\Edit(m')$ is $D \geq \alpha \ell$. 

We have the following theorem.

\BT
For all positive integers $n$ and $\ell =\Omega(\log n)$, assume that Alice and Bob share an $(n,k)$ weak random source $W$ with $k \geq 10 \cdot 2^{3t}\ell$. Then Alice can authenticate $\ell$ bits of message to Bob by using $\mathsf{Protocol}$ $\NAuth(w,m)$. The probability that Eve can successfully change the message to a different string is at most $2^{-\Omega(\ell)}$ and the total number of random bits that Alice and Bob use is $O(\ell)$.

\ET

\begin{thmproof}
The total number of bits that Alice and Bob reveal during the protocol is at most $2 \cdot 2^{3t} t\cdot \ell/t=2 \cdot 2^{3t}\ell$.  At each phase $j$ in Eve's view, where Alice or Bob uses a new chunk of fresh random bits, let $B_{j0}$ stand for the event that conditioned on the fixings of all the strings that Alice and Bob have revealed, the min-entropy of $W$ left is less than $6 \cdot 2^{3t}\ell$. Let $B_0$ stand for the event that there exists a $j$ s.t. $B_{j0}$ happens. Thus by \lemmaref{lem:condition}, we have

\[\Pr[B_{j0}] \leq 2^{-2 \cdot 2^{3t}\ell}, \forall j.\]

By the union bound, 

\[\Pr[B_0] \leq 2t2^{-2 \cdot 2^{3t}\ell}.\]

Now in the event that $B_0$ does not happen, at every phase $j$ the min-entropy of $w$ conditioned on the strings that Alice and Bob have revealed is at least $6 \cdot 2^{3t}\ell$.  Therefore the output of $\Ext(W, X)$ or $\Ext(W,Y)$ is $2^{-\Omega(t)}$-close to uniform even conditioned on $X, Y$ by \theoremref{thm:optext}. 

Let $G$ denote the set of all challenge phases in Eve's view, i.e., $G=\{j$: Phase $j$ is a challenge phase for Eve$\}$. Note that a phase in Eve's view contains at most $2t$ rounds, because each round either Alice or Bob's round number will increase by 1. Thus by \claimref{clm:operation} there are at least $D'/2t \geq D/8t$ bad phases. Now by \lemmaref{lem:challenge} we have 

\[|G| \geq \frac{D}{16t} \geq \frac{\alpha}{16} \cdot \frac{\ell}{t}.\]

Let $E$ denote the event that Eve successfully answered all the challenges. We want to bound $\Pr[E]$ from above.

First let's consider $\Pr[E|\bar{B_0}]$, where $\bar{B_0}$ denotes the event that $B_0$ does not happen. Let $E_j$ stand for the event that Eve successfully answered all the challenges in phase $j$, given all the strings revealed at this time by Alice and Bob and the event $\bar{B_0}$. Thus

\[\Pr[E|\bar{B_0}]=\Pi_{j=1}^{\ell/t} \Pr[E_j|\{E_v:1 \leq v \leq j-1, \bar{B_0}\}].\]

We bound the term $ \Pr[E_j|\{E_v:1 \leq v \leq j-1\}, \bar{B_0}]$ in two cases. First, if $j \notin G$, we simply bound the probability by 1. Second, we bound the probability when $j \in G$.

In this case, since $B_0$ does not happen, conditioned on the fixings of all the previous stings Alice and Bob revealed (and thus all the events $\{E_v:1 \leq v \leq j-1\}$), $W$ has min-entropy at least $6 \cdot 2^{3t}\ell$ and either $\Ext(W, X)$ or $\Ext(W, Y)$ is $2^{-\Omega(t)}$-close to uniform conditioned on the fixings of all previous strings and $X, Y$. Since this is a challenge phase, Eve has to come up with a substring of $\Ext(W,X)$ or $\Ext(W,Y)$. Now let the unfixed strings revealed by Alice and Bob be $s_r$ and let the string that Eve tries to come up with be $s_c$. Then by the definition of a challenge phase $|s_c| \geq |s_r|+2t$ and $S_c$ is $2^{-\Omega(t)}$-close to uniform conditioned on all the fixings. We need to bound the probability of $\Pr[A(S_r)=S_c]$, where $A$ is any (deterministic) algorithm.

We first consider the case where $S_c$ is truly uniform. Now by \lemmaref{lem:condition},

\[\Pr_{S_r}\left[  H_\infty(S_c|(S_r=s_r)) \geq |s_c|-|s_r|-t \right] \geq 1-2^{-t}.\]

That is,
\[\Pr_{S_r}\left[  H_\infty(S_c|(S_r=s_r)) \geq t \right] \geq 1-2^{-t}.\]

Thus 

\[\Pr[A(S_r)=S_c] \leq 2^{-t}+(1-2^{-t}) \cdot 2^{-t} \leq 2^{-t+1}.\]

Now since $S_c$ is $2^{-\Omega(t)}$-close to uniform, we have

\[\Pr[A(S_r)=S_c] \leq 2^{-t+1}+2^{-\Omega(t)}=2^{-\Omega(t)}.\]

Thus we have $\Pr[E_j|\{E_v:1 \leq v \leq j-1\}, \bar{B_0}] \leq 2^{-\Omega(t)}$ when $j \in G$.

Therefore 

\[\Pr[E|\bar{B_0}] \leq (2^{-\Omega(t)})^{|G|} = (2^{-\Omega(t)})^{\frac{\alpha}{16} \cdot \frac{\ell}{t}}=2^{-\Omega(\ell)}.\]

Thus

\[\Pr[E] \leq \Pr[B_0]+\Pr[E|\bar{B_0}] \leq 2t2^{-2 \cdot 2^{3t}\ell}+2^{-\Omega(\ell)}=2^{-\Omega(\ell)}.\]

The total number of random bits that Alice and Bob use is $O(t) \cdot \ell/t=O(\ell)$. The entropy loss of $W$ is at most $2 \cdot 2^{3t}\ell$.
\end{thmproof}

Note that in order for the extractor from \theoremref{thm:optext} to work, we need $O(t) > \log n$. On the other hand we need the entropy of $W$ to be $k \geq 10 \cdot 2^{3t}\ell$. Since $k$ cannot be bigger than $n$, we need $t=O(\log n)$. Thus we can set $t=\Theta(\log n)$ such that $2^{3t}=n^{\gamma}$ for any $0<\gamma<1$. 

In the above analysis we focused on the case where $\ell =\Omega(\log n)$. In the case where $\ell = o(\log n)$, since Alice needs to authenticate $\Theta(\ell+\log n)$ bits to Bob, we can treat it as if $\ell=\Omega(\log n)$ and we get a better security parameter. Thus we have the following theorem.

\BT
For all positive integers $n, \ell$ and every $0<\gamma<\beta<1$, assume that Alice and Bob share an $(n,k)$ weak random source $W$ with $k \geq n^{\beta}$. Then there exists an efficient $(k, \ell)$ interactive authentication protocol such that Alice can authenticate $\Theta(\ell+\log n)$ bits of message to Bob. The total random bits that Alice and Bob use is $O(\ell+\log n)$. The entropy loss of the protocol is $n^{\gamma}(\ell+\log n)$.

\ET

Using the method to convert an authentication protocol to a privacy amplification protocol in \cite{KR09}, \theoremref{thm:authprivacy}, we obtain \theoremref{thm:main1}.

\section{Using Local Weak Ransom Sources} \label{sec:weaksource}
In this section we show how the problem of privacy amplification can be solved when Alice and Bob each only has a local weak random source, instead of truly random bits. As usual we assume that Alice and Bob's weak random sources are independent of each other and independent of the shared weak random source.

\subsection{Non Constructive Results}
First we show that non-constructively, this can be done. In fact, we can essentially reduce the problem to the case where Alice and Bob have local random bits. First we have the following theorem, that can be easily proved by the probabilistic method:

\BT(Two source extractor) \label{thm:text} For all positive integers $n,k$ such that $k > \log n$, there exists a function $\TExt: \bits^n \times \bits^n \to \bits^m$ and $0<\e<1$ such that $m=\Omega(k)$, $\e=2^{-\Omega(k)}$ and if $X, Y$ are two independent $(n,k)$-sources, then

\[ |(X, \TExt(X, Y))-(X, U_m) | \leq \e\]
and
\[ |(Y, \TExt(X, Y))-(Y, U_m) | \leq \e\]
\ET

Now we have the following protocol.

$\mathsf{Protocol}$ $\NExtract(x,y,w)$:
\begin{itemize}
\item Alice has a weak random source $X$, Bob has an independent weak random source $Y$, and they share an independent weak random source $W$. All these sources have min-entropy $k > \polylog(n)$.
\item Let $\TExt$ be the strong two source extractor from \theoremref{thm:text}.

\end{itemize}

\begin{enumerate}
\item Alice and Bob each applies $\TExt$ to his or her own source and $W$.
\item Alice obtains $S_x=\TExt(X, W)$ and Bob obtains $S_y=\TExt(Y,W)$, each outputting $\Omega(k)$ bits.

\end{enumerate}

Now we have the following theorem.

\BT
\[|(S_x, S_y, W)-(U_x, U_y, W) | \leq 2^{-\Omega(k)},\]
where here $(U_x, U_y)$ is the uniform distribution independent of $W$. 
\ET

\begin{thmproof}
By \theoremref{thm:text} and a standard averaging argument, with probability $1-\sqrt{\e}$ over the fixings of $W$, $S_x$ is $\sqrt{\e}$-close to uniform, where $\e=2^{-\Omega(k)}$. Similarly, with probability $1-\sqrt{\e}$ over the fixings of $W$, $S_y$ is $\sqrt{\e}$-close to uniform. Thus with probability $1-2\sqrt{\e}$ over the fixings of $W$, both $S_x$ and $S_y$ are $\sqrt{\e}$-close to uniform. Note that after fixing $W$, $S_x$ is a function of $X$ and $S_y$ is a function of $Y$. Thus they are independent. Therefore with probability $1-2\sqrt{\e}$ over the fixings of $W$, $(S_x, S_y)$ is $2\sqrt{\e}$-close to uniform. Thus we have

\[|(S_x, S_y, W)-(U_x, U_y, W) | \leq 2^{-\Omega(k)}.\]
\end{thmproof}

Now all we need to do is to plug in the non-explicit optimal privacy amplification in \cite{DW09} to obtain \theoremref{thm:main2}.

\subsection{Weak random sources with entropy rate $>1/2$}
Now we study a simple case where Alice and Bob's weak random sources have entropy rate $>1/2$. In this case, we show that we can also reduce the problem to the case where Alice and Bob have local random bits. The reason is that we have strong two-source extractors for such sources, namely Raz's extractor from \theoremref{thm:razweakseed}.

First we have the following protocol.

$\mathsf{Protocol}$ $\ExtractH(x,y,w)$:
\begin{itemize}
\item Alice has a weak random source $X$, Bob has an independent weak random source $Y$, and they share an independent weak random source $W$. Both $X$ and $Y$ have min-entropy $(1/2+\delta) n$ and $W$ has min-entropy $k > \polylog(n)$.  

\item Let $\Raz$ be the strong two source extractor from \theoremref{thm:razweakseed}.

\end{itemize}

\begin{enumerate}
\item Alice and Bob each applies $\Raz$ to his or her own source and $W$.
\item Alice obtains $S_x=\Raz(X, W)$ and Bob obtains $S_y=\Raz(Y,W)$, each outputting $\Omega(k)$ bits.

\end{enumerate}

Now we have the following theorem.

\BT
\[|(S_x, S_y, W)-(U_x, U_y, W) | \leq 2^{-\Omega(k)},\]
where here $(U_x, U_y)$ is the uniform distribution independent of $W$. 
\ET

\begin{thmproof}
Essentially repeat the proof in the previous section.
\end{thmproof}

Again, all we need to do now is to plug in any privacy amplification protocol in \cite{RW03, KR09, DW09, ChandranKOR10} to obtain \theoremref{thm:main3}.

\subsection{Weak random sources with linear min-entropy}
In this section we relax the assumption and only require Alice and Bob have weak random sources with arbitrarily linear min-entropy. More specifically, we assume that Alice and Bob each has a local $(n, \delta n)$ source for some constant $0<\delta<1$. We assume the shared source is an $(n, k)$ source with $k \geq \polylog(n)$. Actually we can also deal with the case where the shared source has linear min-entropy but the local weak sources only have poly logarithmic entropy. This case is quite similar, and thus omitted.  

\subsubsection{The protocol}

Here we give a protocol for Alice and Bob to extract private local random bits. That is, in the end of the protocol, both Alice and Bob obtain local random bits that are close to uniform and independent of the shared weak random source, even in Eve's view. Moreover the shared weak source still has most of its entropy left.

We need the following definition about the slice of a concatenation of strings.

\BD \cite{Rao09} Given $\ell$ strings of length $n$, $x=x_1, \cdots , x_{\ell}$, define $\Slice(x,s)$ to be the string $x' =x'_1, \cdots, x'_{\ell}$ such that for each $i$ $x'_i$ is the prefix of $x_i$ of length $s$. 
\ED

Now we can describe our protocol. In this protocol when a party is authenticating a message to the other party, we do not use the error correcting code. Instead, we just convert the message to a string with a fixed number of 1's. One simple way to do this is map each bit 0 to 01 and map each bit 1 to 10. Thus the number of 1's in the authenticated message is known to both parties before they execute the protocol.

$\mathsf{Protocol}$ $\Extract(x,y,w)$:
\begin{itemize}
\item Alice has a weak random source $X$, Bob has an independent weak random source $Y$, and they share an independent weak random source $W$. Both $X$ and $Y$ have min-entropy $\delta n$ and $W$ has min-entropy $k > \polylog(n)$. 
\item Let $\zuc$ be the somewhere condenser from \theoremref{thm:swcondenser}.
\item Let $\Raz$ be the strong two source extractor from \theoremref{thm:razweakseed}.
\item Let $\SRGExt$ be the two source extractor from \theoremref{thm:srgeneral}.
\item Let $\Ext$ be a strong extractor as in \theoremref{thm:optext}.
\item Let $0<\gamma<1$ be some constant.

\end{itemize}

\begin{enumerate}
\item Alice uses $\zuc$ to convert $X$ into a somewhere rate-$.9$ source $\bar{X}$, with $D$ rows for some constant $D>1$. Similarly Bob also converts $Y$ into a somewhere rate-$.9$ source $\bar{Y}$ with $D$ rows.

\item  Alice applies $\Raz$ to each row of $\bar{X}$ and $W$ and obtains a somewhere random source $SR_x$, with each row outputting $k^{\gamma}$ bits. Similarly Bob also applies $\Raz$ to each row of $\bar{Y}$ and $W$ and obtains a somewhere random source $SR_y$, with each row outputting $k^{\gamma}$ bits.

\item Alice produces 3 strings: $X_1=\Slice(SR_x, c \log n)$, $X_2=\Slice(SR_x, \mu \log k)$ and $X_3=\Slice(SR_x, k^{\beta})$ for some parameters $c>1$, $0<\mu<1$ and $0<\beta<1$ to be chosen later. Bob also produces 3 strings: $Y_1=\Slice(SR_y, c \log n)$, $Y_2=\Slice(SR_y, \mu \log k)$ and $Y_3=\Slice(SR_y, k^{\beta})$.

\item Alice announces $x_1$ to Bob and Bob announces $y_1$ to Alice. Alice then computes $r_y=\Ext(w,y_1)$ and Bob computes $r_x=\Ext(w,x_1)$, where the function $\Ext$ is applied to $w$ and each row of $x_1, y_1$, and each output string has length $k^{\gamma}$.

\item Alice converts $x_2$ to a string $m_x$ with a fixed number of 1's. Let the length of the string be $t$ (note $t=O(\log k)$). Alice then authenticates $m_x$ to Bob by doing the following:

\item Define three set of integers as $C_{1i}=(4D)^{3i-2}c \log n, C_{2i}=(4D)^{3i-1}c \log n, C_{3i}=(4D)^{3i}c \log n$, where $i=1, \cdots, 2t$. 
\item For $i=1$ to $t$ do (authenticate $x_2$ to Bob):

\begin{itemize}
\item If $m_{xi}=0$, Alice sends $(0, \Slice(r_y, C_{1i}))$. Otherwise she sends $(1, \Slice(r_y, C_{2i}))$.

\item Bob receives the message and verifies $\Slice(r_y, C_{1i})=\Slice(\Ext(w,y_1), C_{1i})$ in the $0$ case and $\Slice(r_y, C_{2i})=\Slice(\Ext(w,y_1), C_{2i})$ in the $1$ case. If the verification does not go through, abort. Bob then sends $\Slice(r_x, C_{3i})$ to Alice.

\item Alice receives the message and verifies $\Slice(r_x, C_{3i})=\Slice(\Ext(w,x_1), C_{3i})$.
\end{itemize}

\item When received $t$ bits, Bob verifies that the number of ones in the received string is $\mathsf{wt}(m_x)$; aborts otherwise. Bob recovers $x_2$ from $m_x$.

\item Bob computes $r_3=\SRGExt(y_2, x_2)$, outputting $\Omega(\log k)$ bits. Bob then computes $s_y=\Ext(y_3, r_3)$, outputting $k^{\Omega(1)}$ bits. 

\item Bob converts $r_3$ to a string $m_y$ with a fixed number of 1's. The length of the string is $t'$. Bob then authenticates $m_y$ to Alice by doing the following:

\item For $i=t+1$ to $t+t'$ do (authenticate $r_3$ to Alice): 

\begin{itemize}
\item If $m_{y(i-t)}=0$, Bob sends $(0, \Slice(r_x, C_{1i}))$. Otherwise he sends $(1, \Slice(r_x, C_{2i}))$.

\item Alice receives the message and verifies $\Slice(r_x, C_{1i})=\Slice(\Ext(w,x_1), C_{1i})$ in the $0$ case and $\Slice(r_x, C_{2i})=\Slice(\Ext(w,x_1), C_{2i})$ in the $1$ case. If the verification does not go through, abort. Alice then sends $\Slice(r_y, C_{3i})$ to Bob.

\item Bob receives the message and verifies $\Slice(r_y, C_{3i})=\Slice(\Ext(w,y_1), C_{3i})$.
\end{itemize}

\item When received $t'$ bits, Alice verifies that the number of ones in the received string is $\mathsf{wt}(m_y)$; aborts otherwise. Alice recovers $r_3$ from $m_y$.

\item Alice computes $s_x=\Ext(x_3, r_3)$, outputting $k^{\Omega(1)}$ bits.
\end{enumerate}

\subsubsection{Analysis of the protocol}

We claim that $S_x$ and $S_y$ can now be treated as local private random bits of Alice and Bob. That is , they are close to being independent and uniform and independent of $W$, even in Eve's view. Specifically, we have the following theorem.

\BT Let $V$ denote the transcript of the whole protocol in Eve's view. Then if $S_x \neq \perp$ and $S_y \neq \perp$ (the protocol doesn't abort), we have 

\[|(S_x, S_y, W, V)-(U_x, U_y, W, V) | \leq 1/\poly(k),\]
where here $(U_x, U_y)$ is the uniform distribution independent of $(W, V)$. Moreover, with probability $1-2^{-k^{\Omega(1)}}$ over the fixings of $V=v$, $W$ has min-entropy $k-k^{\Omega(1)}$.
\ET

\begin{thmproof}
Without loss of generality assume that the first row of $\bar{X}$ and the first row of $\bar{Y}$ have entropy rate $0.9$. Let the two rows be $\bar{X}_1$ and $\bar{Y}_1$. Thus by \theoremref{thm:razweakseed} we have

\[|(SR_{x1}, W)-(U_x, W)| = 2^{-\Omega(k)}\] and 

\[|(SR_{y1}, W)-(U_y, W)| = 2^{-\Omega(k)},\]
where $SR_{x1}$ and $SR_{y1}$ stand for the first rows of $SR_x$ and $SR_y$ respectively. Since conditioned on any fixing of $W=w$, $SR_{x1}$ and $SR_{y1}$ are functions of $X$ and $Y$ and are thus independent, we have 

\begin{equation} \label{eqn:SRW}
|(SR_{x1}, SR_{y1}, W)-(U_x, U_y, W) | = 2^{-\Omega(k)}.
\end{equation}

Note that the length of $r_3$ is less than the length of $x_2$. Thus $t' < t$ and therefore the protocol runs for at most $2t=O(\log k)$ rounds. Also in the protocol $\Ext(W, X_{11})$ and $\Ext(W, Y_{11})$ output at most $(4D)^{6t}c \log n=k^{\Omega(1)}\log n$ bits. We choose $\mu$ s.t. this number is at most $k^{\gamma}$, thus we have enough entropy in $W$ for the outputs. Therefore by \equationref{eqn:SRW}, 

\[|(\Ext(W, X_{11}), \Ext(W, Y_{11}))-(U'_x, U'_y) | = 2^{-\Omega(k)}+1/\poly(n)=1/\poly(n).\]

Note that now the random variable that Alice is trying to send to Bob, $X_2$, and the random variables $X_1$, $Y_1$ that have already been revealed, may not be (close to) independent of $(\Ext(W, X_{11}), \Ext(W, Y_{11}))$. We first show in this case the probability that Eve can successfully change a string $x_2$ to a different string is small. To show this, we have the following lemma.

\BL \label{lem:authen}Assume that $(\Ext(W, X_{11}), \Ext(W, Y_{11}))$ is $\e_0$-close to uniform. Let $X_1$ and $Y_1$ be as in the protocol. Let $M$ be any random variable with at most $D \log n$ bits and Alice uses the protocol to authenticate $M$ to Bob. Then the probability that Eve can successfully change a string $m$ to a different string is bounded above by $1/\poly(n)+\e_0$, where the probability is over $M$ and the random variables used to transfer $M$.

\EL

\begin{proof}
Let $\bar{R_x}=\Ext(W, X_1)$, $\bar{R_y}=\Ext(W, Y_1)$ and $\bar{R_{x1}}$, $\bar{R_{y1}}$ be the first rows of $\bar{R_x}$, $\bar{R_y}$ respectively. Thus $\bar{R_{x1}}=\Ext(W, X_{11})$ and $\bar{R_{y1}}=\Ext(W, Y_{11})$. Let $R_x$ and $R_y$ be the actual random variables computed by Bob and Alice respectively. We want to deal with the ideal case where $\bar{R_{x1}}, \bar{R_{y1}}$ is uniform instead of $\e_0$-close to uniform. Note that $(M, X_1, Y_1, \bar{R_x}, \bar{R_y}, R_x, R_y)$ are all the random variables used by Alice to authenticate $M$ to Bob. Thus by \lemmaref{lem:convert} we first construct another distribution $(M', X'_1, Y'_1, \bar{R'_x}, \bar{R'_y}, R'_x, R'_y, \bar{R'_{x1}}, \bar{R'_{y1}})$ where $(\bar{R'_{x1}}, \bar{R'_{y1}})$ is uniform and 

\[|(M, X_1, Y_1, \bar{R_x}, \bar{R_y}, R_x, R_y, \bar{R_{x1}}, \bar{R_{y1}})-(M', X'_1, Y'_1, \bar{R'_x}, \bar{R'_y}, R'_x, R'_y, \bar{R'_{x1}}, \bar{R'_{y1}}) |  \leq \e_0.\]

From now on we will continue the discussion as if $(M, X_1, Y_1, \bar{R_x}, \bar{R_y}, R_x, R_y, \bar{R_{x1}}, \bar{R_{y1}}) = \\ (M', X'_1, Y'_1, \bar{R'_x}, \bar{R'_y}, R'_x, R'_y, \bar{R'_{x1}}, \bar{R'_{y1}})$. We can do this because in the analysis all we use are the sizes of $M', X'_1, Y'_1, \bar{R'_x}, \bar{R'_y}, R'_x, R'_y, \bar{R'_{x1}}, \bar{R'_{y1}}$, which are the same for those of $M', X'_1, Y'_1, \bar{R'_x}, \bar{R'_y}, R'_x, R'_y, \bar{R'_{x1}}, \bar{R'_{y1}}$ by \lemmaref{lem:convert}. Thus the success probability of Eve can only differ by at most $\e_0$.

Now note that the length of $m$ is at most $D \log n$. Thus by \lemmaref{lem:condition} we have

\[\Pr_{M} [H_{\infty}(\Ext(W, X_{11})|M=m)  \geq (4D)^{6t}c \log n-D \log n-D \log n] \geq 1-2^{-D \log n}.\]

That is,

\[\Pr_{M} [H_{\infty}(\Ext(W, X_{11})|M=m)  \geq (4D)^{6t}c \log n-2 D \log n] \geq 1-1/\poly(n). \]

Similarly

\[\Pr_{M} [H_{\infty}(\Ext(W, Y_{11})|M=m)  \geq (4D)^{6t}c \log n-2 D \log n] \geq 1-1/\poly(n). \]

We show that when $m$ is a string s.t. both $(\Ext(W, X_{11})|M=m)$ and $(\Ext(W, Y_{11})|M=m)$ have min-entropy at least $(4D)^{6t}c \log n-2 D \log n$, the success probability that Eve can change $m$ without being detected is $1/\poly(n)$. By the union bound this happens with probability $1-1/\poly(n)$.

To see this, we first prove the following lemma.

\BL In order to change $m$ to a different string, Eve has to come up with at least one challenge.

\EL

\begin{proof}
To change $m$ to a different string, Eve must take a series of operations. We consider two cases.

\begin{itemize}
\item Case 1: The operations that Eve made include insertion or deletion. In this case the first such operation must incur a challenge. To see this, let $j$ be the round right before the insertion or deletion. Thus at the end of round $j$, Alice has announced at most a total of $D C_{2j}+c D \log n = C_{3j}/4+c D \log n$ bits. Similarly Bob has announced at most a total of $D C_{3j}+c D \log n = C_{1(j+1)}/4+c D \log n$ bits. If it's an insertion, Eve has to come up with at least $C_{1(j+1)} =(4D)^{3j+1}c \log n$ random bits to avoid detection, and we see that

\[C_{1(j+1)}-(C_{3j}/4+c D \log n)-(C_{1(j+1)}/4+c D \log n)-D\log n> 4cD \log n.\]

If it's a deletion, then Alice has announced at most a total of $D C_{2(j+1)}+c D \log n = C_{3(j+1)}/4+c D \log n$ bits and Bob has announced a total of $D C_{3j}+c D \log n = C_{1(j+1)}/4+c D \log n$ bits. Eve has to come up with at least $C_{3(j+1)}=(4D)^{3j+3}c \log n$ random bits to avoid detection, and we see that

\[C_{3(j+1)}-(C_{3(j+1)}/4+c D \log n)-(C_{1(j+1)}/4+c D \log n)-D\log n>4c D \log n.\]

\item Case 2: The operations that Eve made do  not include insertion or deletion. In this case, since the number of 1's in the message is known to Bob, Eve must make at least one operation of changing 0 to 1 and at least one operation of changing 1 to 0. Then the operation of changing 0 to 1 will incur a challenge. To see this, let $j$ be the current round(since Eve does not make operations of insertion and deletion, the round number is the same for Alice, Bob and Eve). Thus now Alice has announced a total of $D C_{1j}+c D \log n = C_{2j}/4+c D \log n$ bits while Bob has announced a total of $D C_{3(j-1)}+c D \log n = C_{1j}/4+c D \log n$ bits. Eve has to come up with at least $C_{2j}=(4D)^{3j-1}c \log n$ random bits to avoid detection, and we see that

\[C_{2j}-(C_{2j}/4+c D \log n)-(C_{1j}/4+c D \log n)-D\log n> 4c D \log n.\]

\end{itemize}
\end{proof}

Now let $j$ be the round that Eve has to answer the first challenge. Let $B_j$ stand for the random variable of all the strings that have been revealed by Alice and Bob till now, and let $l_b$ be the length of the string $b_j$. Let $A_j$ denote the random variable that Eve is trying to come up with, and let $l_a$ be the length of the string $a$. Thus we have just shown that $l_a \geq l_b+4 cD\log n$.

Since both $\Ext(W, X_{11})|M=m)$ and $\Ext(W, Y_{11})|M=m)$ have min-entropy at least $(4D)^{6t}c \log n-2 D \log n$, $A$ has min-entropy $l_a-2 D \log n$. Thus by \lemmaref{lem:condition},

\[\Pr_{B}[H_\infty(A|B=b) \geq l_a-2 D \log n-l_b-D \log n] \geq 1-2^{-D \log n}.\]

Thus

\[\Pr_{B}[H_\infty(A|B=b) \geq D \log n] \geq 1-1/\poly(n).\]

Therefore the probability that Eve can successfully change the string is bounded from above by $1/\poly(n)+2^{-D \log n}=1/\poly(n)$.

Thus, going back to the case where $(\Ext(W, X_{11}), \Ext(W, Y_{11}))$ is $\e_0$-close to uniform, the success probability of Eve is bounded from above by $1/\poly(n)+\e_0$.
\end{proof}

Thus the success probability of Eve changing $x_2$ to a different string is bounded from above by $1/\poly(n)+1/\poly(n)=1/\poly(n)$. Note this probability is also over $X_2$. By a standard averaging argument, with probability $1-1/\poly(n)$ over $X_2$, the success probability of Eve changing $x_2$ to a different string is at most $1/\poly(n)$.

Now Bob obtains a random variable $X'_2$. Note that $X'_2$ is not exactly $X_2$ since Eve may be able to change $X_2$ for a probability mass of $\e=1/\poly(n)$.  Assume for now that Bob obtains $X_2$ instead of $X'_2$. Now we fix $W=w$. Note that after this fixing, $X_1, X_2$ are functions of $X$ and $Y_1, Y_2$ are functions of $Y$. By \theoremref{thm:razweakseed}, with probability $1-2^{-\Omega(k)}$ over the fixings of $W=w$, $X_2$ is $2^{-\Omega(k)}$-close to being a somewhere random source, and so is $Y_2$. Moreover $X_2$ and $Y_2$ are independent. Thus by \theoremref{thm:srgeneral}, we have that for a typical fixing of $W=w$,

\begin{equation}\label {eqn:err1}
|(X_2, R_3)-(X_2, U_m) |< \e_1
\end{equation}

and

\begin{equation}\label {eqn:err2}
|(Y_2, R_3)-(Y_2, U_m) |< \e_1,
\end{equation}
where $\e_1=2^{-\Omega(k)}+2^{-\Omega(\log k)}=1/\poly(k)$.

We then further fix $Y_2=y_2$. By \equationref{eqn:err2} with probability $1-\sqrt{\e_1}$ over the fixings of $Y_2=y_2$, $R_3$ is $\sqrt{\e_1}$-close to uniform. Further note that after this fixing $R_3$ is a deterministic function of $X$, and $Y_1$ is a deterministic function of $Y$. Thus we can further fix $Y_1=y_1$ and $R_3$ is still $\sqrt{\e_1}$-close to uniform. Note that $y_1$ has length $cD \log n$ and $Y_3$ has min-entropy $k^{\beta}$. Thus by \lemmaref{lem:condition} we have that with probability $1-1/\poly(n)$ over the fixings of $Y_1=y_1$, $Y_3$ has min-entropy $0.9k^{\beta}$. Thus we have shown that 

\textbf{[Condition 1]} 
With probability $1-2^{-\Omega(k)}-\sqrt{\e_1}-1/\poly(n)=1-1/\poly(k)$ over the fixings of $W=w, Y_2=y_2, Y_1=y_1$, $R_3$ is $\sqrt{\e_1}$-close to uniform, $Y_3$ has min-entropy $k^{\beta}$ and $R_3$ and $Y_3$ are independent.

Now let's consider the case where Bob obtains $X'_2$ instead of $X_2$ and Bob computes $R'_3$ instead of $R_3$. Note that Eve can only change a $\e=1/\poly(n)$ probability mass of $X_2$. For a fixed $W=w, Y_1=y_1$(note that $y_2$ is a slice of $y_1$), let $E_{w,y_1}$ denote the event that Eve changes a $\sqrt{\e}$ probability mass of $X_2|(W=w, Y_1=y_1)$. By a standard averaging argument we have

\[\Pr_{W,Y_1} [E_{w,y_1}] \leq \sqrt{\e}.\]

Now consider a typical fixing of $W=w, Y_1=y_1$ where the event $E_{w,y_1}$ does not happen and \textbf{Condition 1} holds. This happens with probability $1-1/\poly(k)-\sqrt{\e}=1-1/\poly(k)$.  Note since Condition 1 holds, after this fixing $R_3$ and $Y_3$ are independent and $R_3$ is a deterministic function of $X_2$ (and $X$). Now Eve can change a probability mass of $\sqrt{\e}$ here, but all strings revealed by Bob are fixed and $W$ are fixed. Thus whatever Eve does, the resulting $R'_3$ is a function of $X$ and is still independent of $Y_3$. Moreover since Eve can only change a probability mass of $\sqrt{\e}$, $R'_3$ is $\sqrt{\e_1}+\sqrt{\e}=1/\poly(k)$-close to uniform. Therefore we have shown that

\textbf{[Condition 2]} 
With probability $1-1/\poly(k)$ over the fixings of $W=w, Y_2=y_2, Y_1=y_1$, $R'_3$ is $1/\poly(k)$-close to uniform, $Y_3$ has min-entropy $0.9k^{\beta}$ and $R'_3$ and $Y_3$ are independent.

Therefore by the property of the strong extractor $\Ext$, we have

\[|(S_y, R'_3)-(U, R'_3) |< 1/\poly(k). \]

Note that we have fixed $W=w, Y_2=y_2, Y_1=y_1$, and we can now further fix $R'_3=r'_3$. After this fixing $S_y$ is just a function of $Y$ and is independent of $X$. Thus we have fixed all possible information that Eve could know about $Y$ and $S_y$ is still close to uniform. Therefore $S_y$ can be treated as local private random bits of Bob.

Now again by \lemmaref{lem:authen} Bob can authenticate $R'_3$ to Alice such that Eve can only successfully change a probability mass of $\e=1/\poly(n)$ of $R'_3$. Suppose Alice obtains $R''_3$. Now we fix $W=w$ and let $E_w$ stand for the event that Eve changes a $\sqrt{\e}$ probability mass of $X_2|(W=w)$. By a standard averaging argument we have

\[\Pr_{W} [E_w] \leq \sqrt{\e}.\]

Now for a typical fixing of $W=w$ where both $X_2$ and $Y_2$ are close to a somewhere random source and Eve changes less than a $\sqrt{\e}$ probability mass of $X_2|(W=w)$, $X_2$ is a function of $X$, $Y_2$ is a function of $Y$ and are thus independent. By \equationref{eqn:err1} with probability $1-\sqrt{\e_1}$ over the fixings of $X_2=x_2$, $R_3$ is $\sqrt{e_1}$-close to uniform. Thus for a further typical fixing of $X_2=x_2$ where $X_2$ is not changed by Eve and $R_3$ is close to uniform, $R_3$(and $R'_3$) is a function of $Y$ and is independent of $X$. Therefore we can further fix $X_1=x_1$ and $R'_3$ is still close to uniform. Note that $x_1$ has length $cD \log n$ and $X_3$ has min-entropy $k^{\beta}$. Thus by \lemmaref{lem:condition} we have that with probability $1-1/\poly(n)$ over the fixings of $X_1=x_1$, $X_3$ has min-entropy $0.9k^{\beta}$. 

Now Eve can change a $\e=1/\poly(n)$ probability mass of $R'_3$. Let $E_{w,x_2}$ stand for the event that Eve changes a $\sqrt{\e}$ probability mass of $R'_3|(W=w, X_2=x_2)$. By a standard averaging argument we have

\[\Pr_{W, X_2} [E_{w,x_2}] \leq \sqrt{\e}.\]

Thus for a typical fixing of $(W=w, X_2=x_2)$, Eve changes less than $\sqrt{\e}$ probability mass of $R'_3|(W=w, X_2=x_2)$. Since now all strings revealed by Alice and $W$ are fixed, no matter what Eve does, the resulting $R''_3$ is a function of $Y$ and is still independent of $X_3$. Moreover since Eve can only change a probability mass of $\sqrt{\e}$, $R''_3$ is $\sqrt{\e_1}+\sqrt{\e}=1/\poly(k)$-close to uniform. Note the probability of typical fixings of $(W=w, X_2=x_2)$ is at least $1-\sqrt{\e}-\sqrt{\e_1}-\sqrt{\e}-\sqrt{\e}=1-1/\poly(k)$. Therefore we have shown that

\textbf{[Condition 3]} 
With probability $1-1/\poly(k)$ over the fixings of $W=w, X_2=x_2, X_1=x_1$, $R''_3$ is $1/\poly(k)$-close to uniform, $X_3$ has min-entropy $0.9k^{\beta}$ and $R''_3$ and $X_3$ are independent.

Therefore by the property of the strong extractor $\Ext$, we have

\[|(S_x, R''_3)-(U, R''_3) |< 1/\poly(k). \]

Note that we have fixed $W=w, X_2=x_2, X_1=x_1$, and we can now further fix $R''_3=r''_3$. After this fixing $S_x$ is just a function of $X$ and is independent of $Y$, and is thus also independent of $S_y$ (which now is a function of $Y$). Thus we have fixed all possible information that Eve could know about $X$ and $S_x$ is still close to uniform. Therefore $S_x$ can be treated as local private random bits of Bob.

Therefore, we have eventually shown that

\[|(S_x, S_y, X_1, Y_1, W)-(U_x, U_y, X_1, Y_1, W) | \leq 1/\poly(k). \]

Note that now the entire transcript $V$ up till now is a deterministic functions of $W, X_1, Y_1$. Therefore we also have

\[|(S_x, S_y, V, W)-(U_x, U_y, V, W) | \leq 1/\poly(k). \]

Note that the transcript has length at most $k^{\gamma}$. Therefore by \lemmaref{lem:condition} with probability $1-2^{-k^{\Omega(1)}}$ over the fixings of the transcript, $W$ still has min-entropy at least $k-k^{\Omega(1)}$. Thus the theorem is proved.
\end{thmproof}

Now all we need to do is to plug in any privacy amplification protocol in \cite{RW03, KR09, DW09, ChandranKOR10} to obtain \theoremref{thm:main4}.

\section{Conclusions and Open Problems} \label{sec:con}
In this paper we investigated two questions about the local randomness in privacy amplification with an active adversary. The first is what is the minimum number of local random bits needed and the second is whether privacy amplification can be achieved when the two parties only have access to local weak random sources.

For the first question, we showed that $\Theta(\ell +\log n)$ local random bits suffice to achieve security parameter $\ell$, as long as the shared weak random source $W$ has min-entropy $n^{\beta}$ for an arbitrary constant $0<\beta<1$. For the second question, we give positive answers and show that if the local weak random sources have entropy rate $>1/2$, then we can do essentially as good as if we have local uniform random bits. In the case where the local weak random sources have arbitrarrily linear entropy, we show that we can achieve a security parameter of $\Omega(\log k)$, where $k$ is the min-entropy of the shared weak random source $W$.

It is interesting to compare our results and other results in privacy amplification, to the results in the context of randomness extraction. For example, the case where Alice and Bob have access to local random bits can be compared to the construction of seeded extractors. Both problems are known to have optimal solutions non-constructively. However in the case of extractors, we now have constructions that are asymptotically optimal in all parameters \cite{GuruswamiUV07, DvirW08}. In the privacy amplification case, we only have constructions that are optimal in each one of the parameters: \cite{DW09} optimizes the round complexity, \cite{ChandranKOR10} optimizes the entropy loss and our result optimizes the randomness complexity. It is therefore a natural open problem to come up with protocols that are optimal in all these three parameters. Also, our result only works for entropy $n^{\beta}$, thus it would be interesting to construct new protocols that work for entropy as small as $k=\polylog(n)$.

The case where Alice and Bob only have access to local weak random sources can be compared to the construction of 3-source extractors. In fact, a privacy amplification protocol in this case gives a construction of 3-source extractor. Since currently the best known 3-source extractor requires at least one source to have min-entropy $n^{0.9}$ \cite{Rao06}, we do not hope to improve the entropy requirement of our results by much (actually our protocol can also deal with slightly sub-linear entropy) in the near future. Indeed to achieve this goal would require new techniques in constructing extractors for independent sources. However, for a 3-source extractor we can achieve error $2^{-k^{\Omega(1)}}$, while our protocol only achieves $1/\poly(k)$. Thus the natural open problem here is to try to improve the security parameter to $k^{\Omega(1)}$.

It is also interesting to compare our protocols for local weak random sources to the protocols of network extractors \cite{KalaiLRZ08, KalaiLR09}. There the adversary is sort of passive, in the sense that she doesn't change the messages sent between honest parties. Here the adversary is completely active. Thus our results here can be viewed as an extension of \cite{KalaiLRZ08, KalaiLR09} in the two party case. Thus it would be interesting to study network extractors with an active adversary.

\bibliographystyle{alphabetic}

\bibliography{refs}

\appendix

\section{Proof Sketch for \theoremref{thm:srgeneral} } \label{sec:thmsrg}

Here we give a proof sketch for \theoremref{thm:srgeneral}. The algorithm is essentially the same as that in the proof of Theorem 7.5 in \cite{BarakRSW06}, which works by first converting the two sources into a convex combination of two independent \emph{aligned} somewhere random sources, then using the condenser from \cite{Rao06}. The proof is also essentially the same, except for the following differences. First, in that theorem the somewhere random source has $k^{\gamma}$ rows, while here it only has a constant number of rows. Second, in that theorem the extractor outputs $k-k^{\Omega(1)}$ bits with error $2^{-k^{\Omega(1)}}$. Here we only want to output $\Omega(k)$ bits, but the error is $2^{-\Omega(k)}$. In other words, we want the optimal dependence on the error and pay a price at the output length.

The reason for the difference comes from two aspects. First, since the somewhere random source in \cite{BarakRSW06} has $k^{\gamma}$ rows, the slice chosen can only have width $k^{\Omega(1)}$, which results in an error of $2^{-k^{\Omega(1)}}$. Second, the strong seeded extractors used in \cite{BarakRSW06} cannot achieve error $2^{-\Omega(n)}$. We thus adjust the parameters accordingly and use a different construction of seeded extractors to achieve the optimal error.

First, since here the somewhere random source only has a constant number of rows, we can afford to use slices of width $\Omega(k)$ in each step. Second, we use Raz's extractor from \theoremref{thm:razweakseed} as the strong seeded extractor. Raz's extractor is actually a strong 2-source extractor for one source with entropy rate $>1/2$ and another independent weak source. We use it as a strong seed extractor s.t. the random seed of length $\Omega(k)$ is used as the source with entropy rate $>1/2$. Note that the output of Raz's extractor has length $m$ linear in the minimum of the min-entropies of the two sources, and error $2^{-\Omega(m)}$. By induction it's easy to see that now every source in the algorithm has min-entropy $\Omega(k)$. Thus the error is $2^{-\Omega(k)}$ in each step. Since the somewhere random source has only a constant number of rows we only need to condense the sources for a constant number of times. Thus the overall error is $2^{-\Omega(k)}$ and the output length is $\Omega(k)$.

\end{document}